\documentclass[preprint,aps,showpacs]{revtex4}
\usepackage{graphicx,epsf}
\usepackage{lscape}
\usepackage{citesort}
\usepackage[colorlinks,citecolor=blue,urlcolor=red,linkcolor=red]{hyperref}
\begin{document}

\title{{\Large{\bf \sc{ Vertices of the $D_s^{(*)}D^{(*)}K^*$ }}}}

\author{\small
R. Khosravi$^1$ \footnote {e-mail: rezakhosravi @ cc.iut.ac.ir }, M.
Janbazi$^2$ \footnote {e-mail: mehdijanbazi@yahoo.com} }

\affiliation{\emph{$^1$Department of Physics, Isfahan University of
Technology, Isfahan 84156-83111, Iran
\\$^2$Physics Department, Shiraz University, Shiraz 71454, Iran} }

\begin{abstract}
Taking into account the contributions of the quark-quark,
quark-gluon, and gluon-gluon condensate corrections, the strong form
factors and coupling constants of
$D^*_{s0}D^*_0K^*$, $D_sDK^*$, $D^*_sD^*K^*$, $D_{s1}D_1K^*$,
$D_sD^*K^*$, and $D_{s0}D_1K^*$ vertices are investigated  within the three-point
QCD sum rules method, without and with the $SU_{f}(3)$ symmetry.
\end{abstract}


\maketitle

\section{Introduction}
In high energy physics, investigation of meson interactions depends
on information about the proper functional form of the strong form
factors. Among all vertices, the charmed meson ones, which play an
important role in understanding the final state interactions, are
much more significant. Therefore, researchers have concentrated on
computing the strong form factors and coupling constants connected
to these vertices. Until now, the vertices involving charmed mesons
such as $D^* D^* \rho$\cite{MEBracco}, $D^* D \pi$
\cite{FSNavarra,MNielsen}, $D D \rho$\cite{MChiapparini}, $D^* D
\rho$\cite{Rodrigues3}, $D D J/\psi$ \cite{RDMatheus},  $D^* D
J/\psi$ \cite{RRdaSilva}, $D^*D_sK$, $D^*_sD K$, $D_0 D_s K$,
$D_{s0} D K$ \cite{SLWang}, $D^*D^* P$, $D^*D V$, $D D V$
\cite{ZGWang}, $D^* D^* \pi$ \cite{FCarvalho}, $D_s D^* K$, $D_s^* D
K$ \cite{ALozea}, $D D \omega$ \cite{LBHolanda},  $D_s D_s V$,
$D^{*}_s D^{*}_s V$ \cite{KJ,KJ2}, and $D_1D^*\pi, D_1D_0\pi,
D_1D_1\pi$ \cite{Janbazi} have been studied within the framework of
the QCD sum rules.

The effective Lagrangians for the interaction vertices $D_s D K^*$,
$ D_s D^* K^*$, and $D^*_s D^* K^*$ are \cite{BraChia}:
\begin{eqnarray}\label{eq11}
{\cal L}_{D_s D K^*}&=&i g_{D_s D K^*}~{K^*}^\alpha (\bar D_s \partial_\alpha D  -\partial_\alpha \bar D_s D), \nonumber \\
{\cal L}_{D_s D^* K^*}&=&-g_{D_s D^* K^*} \epsilon^{
\alpha\beta \rho \sigma}\partial_{\alpha} D^*_{\beta}(\partial_{\rho} K^{*}_{\sigma} \bar D_s +D_s \partial_{\rho} \bar K^{*}_{\sigma}), \nonumber \\
{\cal L}_{D_s^* D^* K^*}&=& ig_{D_s^* D^* K^*}[{D_s^*}^\mu(\partial_{\mu}{K^*}^\nu \bar D^*_{\nu}-{K^*}^\nu\partial_{\mu}\bar D^*_{\nu})+(\partial_{\mu}{D^*_s}^\nu K^{*}_{\nu}
-{D_s^*}^\nu\partial_{\mu}K^{*}_{\nu}){\bar D}^{*\mu}\nonumber\\&&+ {K^*}^\mu
({D_s^*}^{\nu}\partial_{\mu} \bar D^*_\nu  -
\partial_{\mu}{D^*_s}^\nu  \bar D^*_\nu) ],
\end{eqnarray}
where $g_{D_s D K^*}$, $ g_{D_s D^* K^*}$, and $g_{D^*_s D^* K^*}$
are the strong form factors. From these Lagrangians, the elements
related to the $D_s D K^*$, $D_s D^* K^*$ and $D^*_s D^* K^*$
vertices can be derived in terms of the strong form factors as:
\begin{eqnarray}\label{eq12}
\langle D(p) D_s(p')   |K^*(q, \varepsilon'')   \rangle &=&
-g_{D_s D K^*}(q^2)\times(p^\mu+p'^\mu)\varepsilon''_\mu , \nonumber\\
\langle D^*(p,\varepsilon) D_s(p') |   K^*(q,\varepsilon'') \rangle &=& i
g_{D_s D^* K^*}(q^2)\times \epsilon^{ \alpha\beta \mu \nu}
p'_\alpha q_\beta \varepsilon_\mu(p) \varepsilon''_\nu(q) ,\nonumber\\
\langle D^*(p,\varepsilon) D_s^*(p', \varepsilon') |  K^*(q,\varepsilon'')
\rangle &=& ig_{D^*_s D^*
K^*}(q^2)\times [{(q^\alpha+p'^\alpha)} g^{\mu\nu}- {(q^{\mu}+p^{\mu})} g^{\nu\alpha}+{q}^\nu g^{\alpha\mu}]\nonumber\\&\times&\varepsilon_{\alpha
}(p)\varepsilon'_{\mu}(p')\varepsilon''_{\nu}(q),
\end{eqnarray}
where $q=p-p'$.

In this work, we decide to calculate the strong form factors and
coupling constants associated with the $ D_{s0}^{*} D_{0}^{*}K^*$,
$D_{s} D K^*$, $ D_{s}^{*} D^{*} K^*$, $D_{s1} D_{1} K^*$, $D_{s}
D^{*} K^*$, and $D^*_{s0} D_{1} K^*$ vertices in the frame work of
the three-point QCD sum rules (3PSR).

The plan of the present paper is as follows: In Section II, the
strong form factor calculation of the $D_{s} D K^*$ vertex is
derived in the frame work of the 3PSR;   computing the quark-quark,
quark-gluon and gluon-gluon condensate contributions in the Borel
transform scheme. In Section III, using necessary changes in the
expression obtained for the $g_{D_{s} D K^*}$, the strong form
factors $g_{ D_{s0}^{*} D_{0}^{*}K^*}$, $g_{ D_{s}^{*} D^{*} K^*}$,
$g_{D_{s1} D_{1} K^*}$, $g_{D_{s} D^{*} K^*}$, and $g_{D^*_{s0}
D_{1} K^*}$ are presented.  The next section depicts our numerical
analysis of the strong form factors as well as the coupling
constants, without and with the $SU_{f}(3)$ symmetry.

\section{The STRONG FORM FACTOR OF $D_s D K^*$ vertex  }
\begin{figure}[th]
\begin{center}
\begin{picture}(90,20)
\put(0,-10){ \epsfxsize=9cm \epsfbox{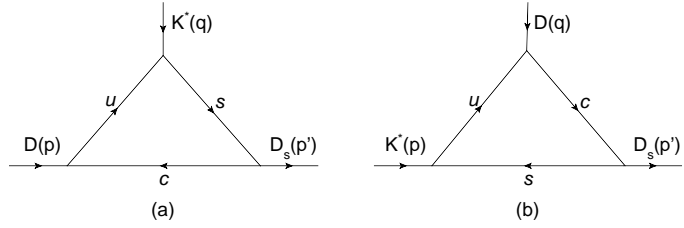} }
\end{picture}
\end{center}
\vspace*{0.5cm} \caption{Perturbative diagrams for off-shell $K^*$
(a) and off-shell $D$  (b).}\label{F1}
\end{figure}

To compute the strong form factor of the $D_s D K^*$ vertex via the
3PSR, we start with the correlation function. When $K^*$ meson is
off-shell, the correlation function is as follows.
\begin{eqnarray}\label{eq21}
\Pi^{K^*}_{\mu}(p, p')&=& i^2 \int d^4x  d^4y e^{i(p'x-py)}\langle 0
|\mathcal{T}\left\{j^{D_s}(x) {j_\mu^{K^*}}^{\dagger}(0)
{j^{D}}^{\dagger}(y)\right\}| 0 \rangle.
\end{eqnarray}
For off-shell charmed meson, the correlation function is:
\begin{eqnarray}\label{eq22}
\Pi^{D}_{\mu}(p, p') &=& i^2 \int d^4x  d^4y e^{i(p'x-py)}\langle 0
|\mathcal{T}\left\{ j^{D_s}(x){j^{D}}^{\dagger}(0){j_{\mu}^{K^*}}^{\dagger}(y)
\right\}| 0 \rangle,
\end{eqnarray}
where $j^{D_{s}}=\bar c \gamma_5 s,~j^{D}=\bar c \gamma_5 u$, and
$j^{K^*}_{\mu}=\bar u \gamma_{\mu} s $ are interpolating currents
with the same quantum numbers of $D_{s},~D$, and $K^*$ mesons,
respectively. Also $\mathcal{T}$ is time ordering product, $p$ and
$p'$ are the four momentum of the initial and final mesons,
respectively as depicted in Fig. \ref{F1}.

The correlation functions in Eqs. (\ref{eq21}) and (\ref{eq22}) are
complex functions of which the real part comprises the computations
of the theoretical part or QCD and imaginary part comprises the
computations of the physical or phenomenological. In the QCD
representation, the correlation function is evaluated in quark-gluon
language like quark-quark, gluon-gluon condensate, etc using the
Wilson operator product expansion (OPE). In the phenomenological
part, the representation is in terms of hadronic degrees of freedom
which is responsible for the introduction of the form factors, decay
constants and masses.

The QCD part of the correlation functions can be calculated  by
expanding it in terms of the OPE, in the deep Euclidean region, as:
\begin{eqnarray}\label{eq23}
\Pi_{\mu}= C^{(0)}_{\mu}{\rm
I}+C^{(3)}_{\mu}\langle0|\bar{\Psi}\Psi|0\rangle+C^{(4)}_{\mu}\langle0|G^{a}_{\rho\nu}G_{a}^{\rho\nu}|0\rangle+C^{(5)}_{\mu}\langle
0| \bar{\Psi} \sigma_{\rho\nu} T^a G_a^{\rho\nu} \Psi| 0\rangle
+...,
\end{eqnarray}
where $C^{(i)}_{\mu}$ are the Wilson coefficients, I is the unit
operator, $\bar{\Psi}$ is the local fermion field operator and
$G^{\rho\nu}$ is the gluon strength tensor. The Wilson coefficient
$C^{(0)}_{\mu}$ is contribution of the perturbative part of the QCD
and the other coefficients are contribution of the non-perturbative
part. The diagrams corresponding to the perturbative (bare loop),
are depicted in Fig. \ref{F1}.

To compute the portion of the perturbative part using the double
dispersion relation for coefficient of the Lorentz structure $p_\mu$
appearing in Eq. (\ref{eq12}), we obtain:
\begin{eqnarray}\label{eq24}
C^{(0)} (p^2, p'^2, q^2)=-\frac{1}{4 \pi^2} \int ds\int ds'\frac{\rho(s, s',q^2)}{(s-p^2)(s'-p'^2)}+\mbox{subtraction  terms},
\end{eqnarray}
where $\rho$ is spectral density. Performing the Fourier
transformation and using the Cutkosky rules, i.e.,
$\frac{1}{p^2-m^2}\rightarrow -2i\pi\delta(p^2-m^2)$, the spectral
densities are calculated for the $p_{\mu}$ structure related to the
$D_s D K^*$ vertex.

$\bullet$ For the off-shell $K^*$ (Fig. $\ref{F1}$ (a)):
\begin{eqnarray*}\label{eq25}
\rho^{K^*}_{ D_s D K^* }&=&6I_0[2 m_cm_s-2m_c^2+\Delta'+ C'_1 (2
m_cm_s-2m_c^2+u) ].
\end{eqnarray*}

$\bullet$ For the off-shell $D$ (Fig. $\ref{F1}$ (b)):
\begin{eqnarray*}\label{eq26}
\rho^{D}_{ D_s D K^*}&=&6I_0[2m_cm_s-2m_s^2+\Delta+ C_1
(2m_cm_s-2m_s^2+2\Delta+ u) ].
\end{eqnarray*}
The explicit expressions of the coefficients in the spectral
densities  are given in Appendix-A.

\begin{figure}[th]
\begin{center}
\begin{picture}(100,20)
\put(0,-25){ \epsfxsize=8cm \epsfbox{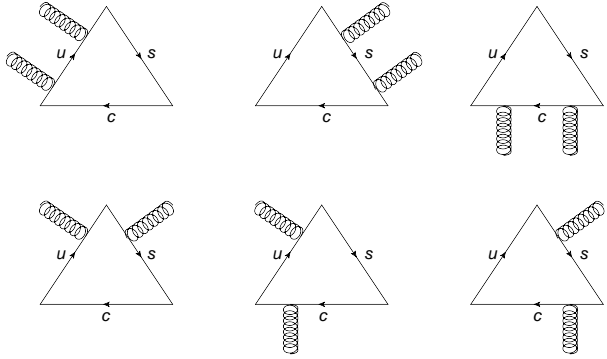} }
\end{picture}
\end{center}
\vspace*{1.5cm} \caption{Non-perturbative diagrams for the off-shell
$K^*$ meson.}\label{F2}
\end{figure}

\begin{figure}[th]
\begin{center}
\begin{picture}(100,00)
\put(0,-20){ \epsfxsize=8cm \epsfbox{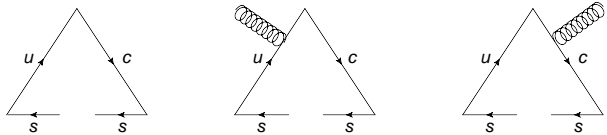} }
\end{picture}
\end{center}
\vspace*{1cm} \caption{Non-perturbative diagrams for the off-shell
$D$ meson.}\label{F3}
\end{figure}

To compute the contribution of the non-perturbative part of the
correlation function for the off-shell $K^*$ meson, six diagrams of
dimension $4$ are considered. These diagrams named gluon-gluon
condensate, shown in Fig. \ref{F2}. In this case the gluon-gluon
diagrams are more important than the other terms in the OPE, since
the heavy $c$ quark is a spectator \cite{Colangelo}.
When $D$ is off-shell, the quark-quark and quark-gluon diagrams of
dimension $3$ and $5$ are more important than the gluon-gluon
condensate, since the light $s$ quark  is a spectator
\cite{Colangelo}. Fig. \ref{F3} shows these diagrams related to the
quark-quark and quark-gluon condensate.

After some straightforward calculations and applying the double
Borel transformations with respect to the $p^2(p^2\rightarrow
M^2_1)$ and $p'^2(p'^2\rightarrow M^2_2)$, where $M_1^2$ and $M_2^2$
are Borel parameters, the following results are obtained for the
non-perturbative contributions corresponding to Fig. \ref{F2} and
Fig. \ref{F3}, respectively:
\begin{eqnarray}\label{eq27}
C^{(4)}=
i\left\langle\frac{\alpha_s}{\pi}G^2\right\rangle\frac{C_{D_sDK^*}^{K^*}}{12},~~~~~
C^{(3)}+C^{(5)}=\langle s\bar s\rangle\frac{C_{D_sDK^*}^{D}}{12},
\end{eqnarray}
where the explicit expressions for $C_{D_sDK^*}^{K^*}$ and
$C_{D_sDK^*}^{D}$  are given in appendix-B. It should be noted that
to obtain the gluon-gluon condensate contributions, we will follow
the same procedure as stated in \cite{Kiselev}.

In order to calculate the phenomenological part of the correlation
functions in Eqs. (\ref{eq21}) and (\ref{eq22}), three complete sets
of intermediate states with the same quantum number should be
inserted in these equations. Performing the Fourier transformation,
for the phenomenological parts, we have:
\begin{eqnarray}\label{eq28}
\Pi^{K^*}_{\mu}&=&\frac{\langle 0|j^{D_s}|D_s(p')\rangle\langle
0|j^{D }|D(p)\rangle\langle D_s(p')D(p)|K^*(q,\epsilon)\rangle
\langle K^*
(q,\epsilon)|j^{K^*}_{\mu}|0\rangle}{(p^2-m^2_{D})(p'^2-m^2_{D_s})(q^2-m^2_{K^*})}
\nonumber\\&&+\mbox{higher and continuum states,}\nonumber\\
\Pi^{D}_{\mu}&=&\frac{\langle 0|j^{D_s}|D_s(p')\rangle\langle
0|j_\mu^{K^*} |K^*(p,\epsilon)\rangle\langle
 D_s(p')K^*(p,\epsilon)|D(q) \rangle \langle D
(q)|j^{D}|0\rangle}{(p^2-m^2_{K^*})(p'^2-m^2_{D_s})(q^2-m^2_{D})}
\nonumber\\&&+\mbox{higher and continuum states.}
\end{eqnarray}
The matrix elements $\langle 0 | j_{\mu}^{K^*} | K^*(q,\varepsilon) \rangle$, and  $\langle 0 | j^{D_{(s)}} | D_{(s)}(p) \rangle$  are defined as:
\begin{eqnarray}\label{eq29}
\langle 0 | j_{\mu}^{K^*} | K^*(q,\varepsilon) \rangle &=& m_{K^*}
f_{K^*} \varepsilon_{\mu}(q), \nonumber \\
\langle 0 | j^{D_{(s)}} | D_{(s)}(p) \rangle &=& \frac{ m^2_{D_{(s)}} f_{D_{(s)}}}{m_c+m_{u(s)}},
\end{eqnarray}
where $m_{K^*}$, $m_{D_{(s)}}$, $f_{K^*}$,  and  $f_{D_{(s)}}$  are the masses and decay constants of mesons $K^*$  and $D_{(s)}$, respectively.
$\varepsilon_\mu$ is the polarization vector
of the vector  meson $K^*$.

Inserting Eqs. (\ref{eq12}) and (\ref{eq29}) in Eq. (\ref{eq28}) and
after some calculations, we obtain $\Pi_{\mu}^{K^*}$ and
$\Pi_{\mu}^{D} $ in terms of strong form factors $g^{K^*}_{D_s D K^*
}$ and $g^{D}_{D_s D K^* }$ as:
\begin{eqnarray}\label{eq210}
\Pi_{\mu}^{K^*} &=& -g^{K^*}_{D_s D K^* }(q^2)\frac{ m_{K^*}m_{D}^2
m_{D_s}^2 f_{K^*}f_{D} f_{D_s}
}{m_c(m_c+m_s)(p^2-m^2_{D})(p'^2-m^2_{D_s})(q^2-m^2_{K^*})}p_{\mu}+\mbox{...}\nonumber\\
\Pi_{\mu}^{D}  &=&-g^{D}_{ D_s D K^* }(q^2)\frac{m_{D_s}^2m_D^2
f_{K^*}f_{D}f_{D_s}(m_{D}^2+m_{K^*}^2-q^2)}{m_c(m_c+m_s)m_{K^*}
(p^2-m_{K^*}^2)(p'^2-m_{D_s}^2)(q^2-m_{D}^2)}p_{\mu}+\mbox{...}
\end{eqnarray}

The strong form factors are calculated by equating two
representations of the correlation function and applying the Borel
transformations with respect to the $p^2(p^2\rightarrow M^2_1)$ and
$p'^2(p'^2\rightarrow M^2_2)$ on the phenomenological as well as the
perturbative and nonperturbative parts of the correlation function
in order to suppress the contributions of the higher states and
continuum. The equations for the strong form factors are obtained as
follows:
\begin{eqnarray}\label{eq212}
g^{K^*}_{D_sDK^* }(q^2)&=&\Lambda^{K^*}_{D_sDK^*}
\left\{-\frac{1}{4\pi^2}\int^{s_0^{D_s}}_{(m_c+m_s)^2}ds'
\int^{s_0^D}_{s_1} ds \rho^{K^*}_{DD_sK^*}(s,s',q^2)
e^{-\frac{s}{M_1^2}}
e^{-\frac{s'}{M_2^2}} \right.\nonumber \\
&-&\left. iM^{2}_{1}M^{2}_{2} \left \langle \frac{\alpha_s}{\pi} G^2
\right\rangle \frac{C_{D_sDK^*}^{K^*}}{12} \right\},\nonumber\\
g^{D}_{D_sDK^*}(q^2)&=&  \Lambda^{D}_{D_sDK^*}
\left\{-\frac{1}{4\pi^2}\int^{s^{D_s}_0}_{(m_c +m_s)^2}ds'
\int^{s^{K^*}_0}_{s_2} ds \rho^{D}_{ DD_sK^*}(s,s',q^2)
e^{-\frac{s}{M_1^2}} e^{-\frac{s'}{M_2^2}} \right. \nonumber \\
&+& \left.  M_1^2 M_2^2~ \langle s\bar s
\rangle\frac{C^{D}_{D_sDK^*}}{12} \right\},
\end{eqnarray}

where $s^{K^*}_0$ and $s^{D(D_s)}_0$ are the continuum thresholds in
$K^*$ and $D(D_s)$ mesons, respectively. $s_1$ and $s_2$ are the
lower limits of the integrals over $s$ as
\begin{eqnarray*}\label{eq225}
s_{1}=\frac{m_c^2(m_c^2-s'+q^2)}{m_{c}^{2}-s'}\,,\quad\quad\quad
s_{2}=\frac{m_s^2(m_s^2-s'+q^2)}{m_{s}^{2}-s'}\,.
\end{eqnarray*}
Also $\Lambda^{K^*}_{D_sDK^*}$ and $\Lambda^{D}_{D_sDK^*}$  are
defined as:
\begin{eqnarray}\label{eq213}
\Lambda^{K^*}_{D_sDK^*}&=&-\frac{m_c(m_c+m_s)(q^2-m_{K^*}^{2})}{m_{K^*} m_{D}^2m_{D_s}^2
f_{K^*}f_{D}f_{D_s}}~ e^{\frac{m_{D}^2}{M_1^2}}e^{\frac{m_{D_s}^2}{M_2^2}},\nonumber\\
\Lambda^{D}_{D_sDK^*} &=&-\frac{m_c(m_c+m_s)m_{K^*}(q^2-m_{D}^{2})}{
m_{D}^2m_{D_s}^2 f_{K^*}f_{D} f_{D_s}(m_{D_s}^2+m_{K^*}^2-q^2)}
~e^{\frac{m_{K^*}^2}{M_1^2}} e^{\frac{m_{D_s}^2}{M_2^2}}.
\end{eqnarray}

\section{OTHER VERTICES AND STRONG FORM FACTORS}
Following  the previous steps in section II, phrases similar  to Eq. (\ref{eq212}) can be obtained for the strong form factors of the $ D_{s0}^{*} D_{0}^{*}K^*$,
$D_{s}^{*} D^{*} K^*$, $D_{s1} D_{1} K^*$, $D_{s} D^{*} K^*$, and
$D^*_{s0} D_{1} K^*$ vertices  via the 3PSR. For this purpose, the appropriate terms of $\Lambda$, the spectral density,
and quark-gluon condensate should be replaced in Eq. (\ref{eq212}).
proper expressions for $\Lambda$, the spectral density, and quark-gluon condensate, related to
the strong form factors $g_{ D_{s0}^{*} D_{0}^{*}K^*}$,
$g_{D_{s}^{*} D^{*} K^*}$, $g_{D_{s1} D_{1} K^*}$, $g_{D_{s} D^{*} K^*}$, and
$g_{D^*_{s0} D_{1} K^*}$, have been listed in the Tables \ref{T331}-\ref{T332} and Appendix-B, respectively.
\begin{table}[th]
\caption{Expressions for the coefficient $\Lambda$ related to
the strong form factors $g_{ D_{s0}^{*} D_{0}^{*}K^*}$,
$g_{D_{s}^{*} D^{*} K^*}$, $g_{D_{s1} D_{1} K^*}$, $g_{D_{s} D^{*} K^*}$, and
$g_{D^*_{s0} D_{1} K^*}$.}
\label{T331}
\begin{ruledtabular}
\begin{tabular}{cc}
Coefficient($\Lambda $) & Expression
\\ \hline
$\Lambda^{D_0^*}_{D^{*}_{s0} D_0^* K^* }$&$ -\frac{m_{K^*}(q^2-m_{D_0^*}^{2})}{
m_{D^*_{s0}}m_{D_0^*} f_{K^*}f_{D^*_{s0}}f_{D_0^*} (m_{D^*_{s0}}^2+m_{K^*}^2-q^2)}~e^{\frac{m_{K^*}^2}{M_1^2}} e^{\frac{m_{D^{*}_{s0}}^2}{M_2^2}}$
\\
$\Lambda^{K^*}_{D^{*}_{s0} D_0^* K^* }$&$
-\frac{(q^2-m_{K^*}^{2})}{m_{K^*} m_{D^*_{s0}}m_{D_0^*}
f_{K^*}f_{D^*_{s0}}f_{D_0^*}}~ e^{\frac{m_{D_0^*}^2}{M_1^2}}e^{\frac{m_{D^{*}_{s0}}^2}{M_2^2}}$
\\
$\Lambda^{D^*}_{D^{*}_{s} D^* K^* }$&$ -\frac{2 m_{D_s^*}(q^2-m_{D^*}^{2})}{ m_{K^*}
f_{K^*}f_{D^*}f_{D_s^*} (3m_{D^*}^2+m_{K^*}^2-q^2)}~e^{\frac{m_{K^*}^2}{M_1^2}} e^{\frac{m_{D^*_s}^2}{M_2^2}}$
\\
$\Lambda^{K^*}_{D^{*}_{s} D^* K^* }$&$ -\frac{2 m_{D_s^*} (q^2-m_{K^*}^{2})}{
m_{K^*} f_{K^*}f_{D^*}f_{D_s^*} (3m_{D_s^*}^2+m_{D^*}^2-q^2)}~ e^{\frac{m_{D^*}^2}{M_1^2}}e^{\frac{m_{D^*_s}^2}{M_2^2}}$
\\
$\Lambda^{D_1}_{D^{*}_{s1} {D}_{1} {K}^{*} }$&$ -\frac{2m_{D_{s1}}(q^2-m_{D_1}^{2})}{ m_{K^*}
f_{K^*}f_{D_1}f_{D_s^*} (3m_{D_{s1}}^2+m_{K^*}^2-q^2)}~e^{\frac{m_{K^*}^2}{M_1^2}} e^{\frac{m_{D^{*}_{s1}}^2}{M_2^2}}$
\\
$\Lambda^{K^*}_{D^{*}_{s1} D_1 K^* }$&$ -\frac{2m_{D_{s1}}(q^2-m_{K^*}^{2})}{
m_{K^*} f_{K^*}f_{D_1}f_{D_{s1}^*} (3m_{D_{s1}^*}^2+m_{D_1}^2-q^2)}~ e^{\frac{m_{D_1}^2}{M_1^2}}e^{\frac{m_{D^{*}_{s1}}^2}{M_2^2}}$
\\
$\Lambda^{D^{*}}_{D_{s}D^{*}K^*}$&$
-\frac{(m_c+m_s)(q^2-m^2_{D^*})}{m_{K^*}m_{D^{*}}m^{2}_{D_{s}}
f_{K^*}f_{D^{*}}f_{D_{s}}}~e^{\frac{m_{K^*}^2}{M_1^2}} e^{\frac{m_{D_s}^2}{M_2^2}}$
\\
$\Lambda^{K^*}_{D_{s}D^{*}K^*}$&$
-\frac{(m_c+m_s)(q^2-m^2_{K^*})}{m_{K^*}m_{D^{*}}m^{2}_{D_{s}}
f_{K^*}f_{D^{*}}f_{D_{s}}}~ e^{\frac{m_{D^*}^2}{M_1^2}}e^{\frac{m_{D_s}^2}{M_2^2}}$
\\
$\Lambda^{D_{1}}_{D_{s0}^{*}D_{1}K^*}$&$
-\frac{(q^2-m^2_{D_{1}})}{m_{K^*}m_{D_{1}}m_{D_{s0}^{*}}
f_{K^*}f_{D_{1}}f_{D_{s0}^{*}}}~e^{\frac{m_{K^*}^2}{M_1^2}} e^{\frac{m_{D_{s0}^{*}}^2}{M_2^2}}$
\\
$\Lambda^{K^*}_{D_{s0}^{*}D_{1}K^* }$&$
-\frac{(q^2-m^2_{K^*})}{m_{K^*}m_{D_{1}}m_{D_{s0}^{*}}
f_{K^*}f_{D_{1}}f_{D_{s0}^{*}}}~ e^{\frac{m_{D_1}^2}{M_1^2}}e^{\frac{m_{D_{s0}^{*}}^2}{M_2^2}}$
\end{tabular}
\end{ruledtabular}
\end{table}
\begin{table}[th]
\caption{Spectral density expressions connected to
the strong form factors $g_{ D_{s0}^{*} D_{0}^{*}K^*}$,
$g_{D_{s}^{*} D^{*} K^*}$, $g_{D_{s1} D_{1} K^*}$, $g_{D_{s} D^{*} K^*}$, and
$g_{D^*_{s0} D_{1} K^*}$.}
\label{T332}
\begin{ruledtabular}
\begin{tabular}{cc}
Spectral Density ($\rho$)& Expression
\\ \hline
$\rho^{D^*_0}_{ D^*_{s0} D^*_0 K^*}$&$6I_0[2m_cm_s+2m_s^2-\Delta+ C_1 (2m_cm_s+2m_s^2-2\Delta+ u) ]$\\
$\rho^{K^*}_{ D^*_{s0} D^*_0 K^* }$&$6I_0[2 m_cm_s+2m_c^2-\Delta'+ C'_1 (2 m_cm_s+2m_c^2-u) ]$\\
$\rho^{D^*}_{ D^*_s D^* K^*}$&$3I_0[3m_s^2-2m_cm_s-s-\Delta+4 A+(C_1-C_2)(2u+2m_cm_s-2s)
-8(E_1- E_2)]$\\
$\rho^{K^*}_{ D^*_s D^* K^* }$&$3I_0[3m_c^2-2m_cm_s-s-\Delta'+4 A'+2(C'_1-C'_2)(u+2m_cm_s-2s)
-8(E'_1- E'_2)]$\\
$\rho^{D_1}_{ D^*_{s1} D_1 K^*}$&$3I_0[3m_s^2+2m_cm_s-s-\Delta+4 A+(C_1-C_2)(2u-2m_cm_s-2s)
-8(E_1- E_2)]$\\
$\rho^{K^*}_{ D^*_{s1} D_1 K^* }$&$3I_0[3m_c^2+2m_cm_s-s-\Delta'+4 A'+2(C'_1-C'_2)(u-2m_cm_s-2s)
-8(E'_1- E'_2)]$\\
$\rho^{D^*}_{ D_s D^* K^*}$&$12I_0[C_1m_s+C_2(m_s-m_c)+m_s]$\\
$\rho^{K^*}_{ D_s D^* K^* }$&$-12I_0[C'_1m_c+C'_2(m_c-m_s)+m_c]$\\
$\rho^{D_1}_{ D^*_{s0} D_1 K^*}$&$12I_0[C_1m_s+C_2(m_s+m_c)+m_s]$\\
$\rho^{K^*}_{ D^*_{s0} D_1 K^*}$&$-12I_0[C'_1m_c+C'_2(m_c+m_s)+m_c]$\\
\end{tabular}
\end{ruledtabular}
\end{table}
\clearpage
\section{NUMERICAL ANALYSIS}
In this section, the strong form factors, and coupling
constants for the $ D_{s0}^{*} D_{0}^{*}K^*$, $D_{s} D K^*$, $
D_{s}^{*} D^{*} K^*$, $D_{s1} D_{1} K^*$, $D_{s} D^{*} K^*$, and
$D^*_{s0} D_{1} K^*$ vertices are considered. For this aim,  the values of  quark and
meson masses are chosen as: $m_s = 0.14\pm0.01~\rm GeV$,
$m_{K^*}=0.89~ \rm GeV$, $m_{D^{*}_{s0}}=2.32~ \rm GeV$,
$m_{D_{s}}=1.97~\rm GeV$, $m_{D^{*}_{s}}=2.11~\rm GeV$,
$m_{D_{s1}}=2.46~\rm GeV$, $m_{D^*_{0}}=2.40~\rm GeV$,
$m_{D}=1.87~\rm GeV$, $m_{D^{*}}=2.01~\rm GeV$ and
$m_{D_{1}}=2.42~\rm GeV$ \cite{PDG2012}. Also the leptonic decay
constants for these vertexes are presented in Table \ref{T31}.
\begin{table}[h]
\caption{The leptonic decay constants in $\rm MeV$. }\label{T31}
\begin{ruledtabular}
\begin{tabular}{cccccccccc}
$f_{K^*}$ \cite{PDG2012}& $f_{D^{*}_{s0}}$ \cite{Colang}& $f_{D^{*}_{0}}$\cite{THuang} & $f_{D_{s}}$ \cite{Artuso}& $f_{D}$ \cite{ZGWang}& $f_{D^{*}_{s}}$ \cite{Colang}& $f_{D^{*}}$ \cite{GLWang} & $f_{D_{s1}}$ \cite{Thoma}& $f_{D_{1}}$ \cite{Bazavov}\\
\hline $220\pm5$&  $230\pm 20$     &$334\pm 9$ & $294\pm 27$&$223\pm 17$ &$266\pm 32$
&$340\pm 12$ &$225\pm 20$&$219\pm 11$
\end{tabular}
\end{ruledtabular}
\end{table}

There are four auxiliary parameters containing the Borel mass
parameters $M_1$ and $M_2$ and continuum thresholds $s^{K^*}_{0}$
and $s_{0}^{D(D_s)}$ in Eq. (\ref{eq212}). The strong form factors
and coupling constants are the physical quantities and should be
independent of them. However the continuum thresholds are not
completely arbitrary; these are related to the energy
of the first exited state. The values of the continuum
thresholds are taken to be $s^{K^*}_{0}=(m_{K^*}+\delta)^2$ and
$s_{0}^{D(D_s)}=(m_{D(D_s)}+\delta)^2$. We use $0.4 ~\rm GeV\leq
\delta \leq0.6~\rm \rm GeV$ in $Q^2=1~\rm GeV^2$, where $Q^2=-q^2$
\cite{FSNavarra,MNielsen,MEBracco}.

Our results should be almost insensitive to the intervals of the Borel  parameters.
On the other hand, the intervals of the Borel mass parameters must suppress the higher states, continuum
and contributions of the highest-order operators. In other words,
the sum rules for the strong form factors must converge. In this work, the following relations between the
Borel masses $M_1$ and $M_2$ are used.

$\bullet$ For the off-shell $K^*$ meson:
\begin{eqnarray}\label{eq31}
\frac{M_1^2}{M_2^2}=\frac{m_{D}^2}{m_{D_s}^2}.
\end{eqnarray}

$\bullet$ For the  off-shell $D$ meson:
\begin{eqnarray}\label{eq32}
\frac{M_1^2}{M_2^2}=\frac{m_{K^*}^2}{m_{D_s}^2-m_c^2}.
\end{eqnarray}

\begin{figure}[th]
\vspace{0cm}
\includegraphics[width=7cm,height=4cm]{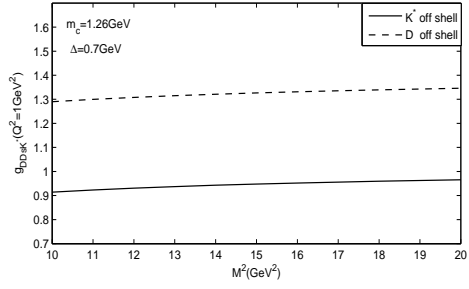}
\hspace{-0.5cm}
\caption{The strong form factors $g_{D_s D K^*}$ as
functions of the Borel mass parameter $M^2$.}\label{F31}
\end{figure}

So, only one independent Borel mass parameter, $M$ is obtained
according to these relations between the $M_1$ and $M_2$. We found a
good stability of the sum rule in the interval $10~\rm GeV^2\leq M^2
\leq20~\rm GeV^2$ for all vertices. For instance, the dependence of
the strong form factor, $g_{D_s D K^*}$ on Borel mass parameter,
$M^2$ in $Q^2=1$ is shown in Fig. \ref{F31}.

To extend the $Q^2$ dependence of the strong form
factors to the full physical region, where the sum rule results are not valid,
we find that the sum rules predictions for the form factors in Eq. (\ref{eq212})  are well fitted
to the following function:
\begin{eqnarray}\label{eq33}
g(Q^2)=A~e^{-Q^2/B}.
\end{eqnarray}

The values of the parameters $A$ and $B$ are given for two sets in Table
\ref{T32}. set I: $m_c=1.26~ \rm GeV$ and set II: $m_c= 1.47~ \rm GeV$.
\begin{table}
\caption{Parameters appearing in the fit functions for the
$D^*_{s0}D^*_0K^*$, $D_sDK^*$,  $D^*_sD^*K^*$, $D_{s1}D_1K^*$,
$D_sD^*K^*$, and $D_{s0}D_1K^*$, vertices for various $m_c$ and
$\delta$, where  $\delta_1=0.4 ~\rm GeV, \delta_2=0.7 ~\rm GeV$ and
$\delta_3=1.0 ~\rm GeV$.}\label{T32}
\begin{ruledtabular}
\begin{tabular}{ccccccccc}
&$\mbox{set I}$&$$$$&$$&$$&$$&$$&$\mbox{set II}$&\\
\hline
$\mbox{Form factor}$&$A(\delta_1)$&$B(\delta_1)$&$A(\delta_2)$&$B(\delta_2)$&$A(\delta_3)$&$B(\delta_3)$&$A(\delta_2)$&$B(\delta_2)$\\
$g^{K^*}_{D^*_{s0}D_0K^*}(Q^2)$&5.32&3.24&6.03&4.61&6.96&7.57&4.46&2.21   \\
$g^{D_0}_{D^*_{s0}D_0K^*}(Q^2)$&5.33&26.60&5.78&27.16&6.56&28.06&5.22&26.13    \\
$g^{K^*}_{D_sDK^*}(Q^2)$&2.22&2.77&2.58&3.07&3.19&4.20&2.21&1.52 \\
$g^{D}_{D_sDK^*}(Q^2)$&2.71&30.28&3.09&30.21&3.49&30.26&3.22&33.93 \\
$g^{K^*}_{D^*_sD^*K^*}(Q^2)$&4.36&191.56&4.91&199.52&5.67&212.43&4.63&438.03 \\
$g^{D^*}_{D^*_sD^*K^*}(Q^2)$&3.58&52.74&4.42&49.58&5.06&47.41&4.20&53.70 \\
$g^{K^*}_{D_{s1}D_1K^*}(Q^2)$&3.78&18.13&4.26&26.95&4.64&40.86&4.52&8.49 \\
$g^{D_1}_{D_{s1}D_1K^*}(Q^2)$&2.76&18.44&3.18&21.05&3.89&26.76&3.31&16.26 \\
$g^{K^*}_{D_sD^*K^*}(Q^2)$&3.98&13.20&4.15&7.34&4.39&6.07&4.26&5.32 \\
$g^{D^*}_{D_sD^*K^*}(Q^2)$&3.98&39.11&4.26&34.29&4.56&28.46&4.42&30.64   \\
$g^{K^*}_{D^*_{s0}D_1K^*}(Q^2)$&7.32&14.57&8.31&32.56&9.09&56.18&8.58&15.17\\
$g^{D_1}_{D^*_{s0}D_1K^*}(Q^2)$&5.33&17.06&5.63&15.60&5.75&12.96&6.35&15.58 \\
\end{tabular}
\end{ruledtabular}
\end{table}

\begin{figure}[th]
\vspace{0cm}
\includegraphics[width=7cm,height=4cm]{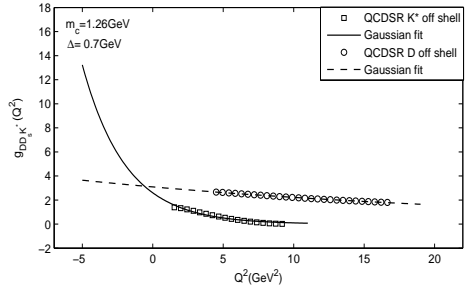}
\caption{The strong form factors $g^{D}_{D_sDK^*}$
and $g^{K^*}_{D_sDK^*}$ on $Q^2$. }\label{F32}
\end{figure}

The dependence of the strong form factors $g^{D}_{D_sDK^*}(Q^2)$ and
$g^{K^*}_{D_sDK^*}(Q^2)$ in $Q^2$ are shown in Fig. \ref{F32}. In
this figure, the small circles and boxes correspond to the form
factors via the 3PSR calculation. As it is seen, the form factors and
their fit functions coincide together, well.

The value of the strong form factors at $Q^2 = -m_m^2$, where $m_m$ is the mass of the off-shell meson, is defined as coupling constant.
Coupling constant results of the vertices,  ${D^*_{s0}D^*_0K^*}$,
${D_sDK^*}$, ${D^*_sD^*K^*}$, ${D_{s1}D_1K^*}$,
${D_sD^*K^*}$, and ${D_{s0}D_1K^*}$ are presented in Table \ref{T33}.
\begin{table}[th]
\caption{The  coupling constant of the vertices  ${D^*_{s0}D^*_0K^*}$,
${D_sDK^*}$, ${D^*_sD^*K^*}$, ${D_{s1}D_1K^*}$,
${D_sD^*K^*}$, and ${D_{s0}D_1K^*}$, in $\rm GeV^{-1}$ for
various $m_c$. }\label{T33}
\begin{ruledtabular}
\begin{tabular}{cccccc}
&$\mbox{set I}$&$$&$\mbox{set II}$&$$\\
$g$&$\mbox{off-shell charmed }$&$\mbox{off-shell $K^*$ }$&$\mbox{off-shell charmed }$&$\mbox{off-shell $K^*$ }$\\
\hline
$g_{D^*_{s0}D^*_0K^*}$&$7.09\pm0.91$&$7.16\pm0.71$&$6.46\pm0.88$&$6.39\pm0.81$\\
$g_{D_sDK^*}$&$3.47\pm0.45$&$3.37\pm0.48$&$3.57\pm0.47$&$3.72\pm0.42$\\
$g_{D^*_sD^*K^*}$&$4.79\pm0.72$&$4.93\pm0.74$&$4.53\pm0.66$&$4.63\pm0.76$\\
$g_{D_{s1}D_1K^*}$&$4.29\pm1.34$&$4.42\pm0.90$&$4.05\pm1.13$&$3.88\pm1.06$\\
$g_{D_sD^*K^*}$&$4.80\pm0.45$&$4.62\pm0.38$&$5.04\pm0.37$&$4.95\pm0.41$\\
$g_{D^*_{s0}D_1K^*}$&$8.19\pm0.81$&$8.51\pm0.78$&$9.25\pm0.76$&$9.04\pm0.81$
\end{tabular}
\end{ruledtabular}
\end{table}
The errors are estimated by the variation of the Borel parameter,
the variation of the continuum thresholds, the leptonic decay
constants and uncertainties in the values of the other input
parameters. It should be noted that the main uncertainty comes from the continuum thresholds
and the decay constants.

Table \ref{T34} shows a comparison between our results with the values predicted by the light-cone sum rules
(LCSR) method. The results of Ref. \cite{Wang}
have been rescaled according to the strong form factor definitions in
Eq. (\ref{eq12}). It should be reminded that the average value of two coupling constants,
$g^{D}_{D_sDK^*}$ and $g^{K^*}_{D_sDK^*}$ in set I in Table \ref{T33} , are presented in Table \ref{T34}. As seen, our values are in the reasonable agreement with those of the LCSR.
\begin{table}[th]
\caption{Values of the strong coupling constant using the 3PSR (ours) and LCSR
approaches, in
$\rm GeV^{-1}$.} \label{T34}
\begin{ruledtabular}
\begin{tabular}{ccc}
$g$ & Ours &LCSR \cite{Wang}
\\ \hline
$g_{D_sDK^*}$ &$3.42\pm0.44$& $3.22\pm0.66$\\
$g_{D_sD^*K^*}$ &$4.71\pm0.39$& $4.04\pm0.80$
\end{tabular}
\end{ruledtabular}
\end{table}

In order to investigate the strong coupling constant values via the
$SU_{f}(3)$ symmetry, the mass of the $s$ quark are ignored in all
equations. In view of the
$SU_{f}(3)$ symmetry, the values of the parameters $A$ and $B$ for
the $g_{D^*_{s0}D^*_0K^*}$, $g_{D_sDK^*}$, $g_{D^*_sD^*K^*}$,
$g_{D_{s1}D_1K^*}$, $g_{D_sD^*K^*}$, and $g_{D_{s0}D_1K^*}$ vertices
in $m_c=1.26~\rm GeV$ and $\delta=0.7~\rm GeV$ are given in Table
\ref{T35}.
\begin{table}[th]
\caption{Parameters appearing in the fit functions for the
$g_{D^*_{s0}D^*_0K^*}$, $g_{D_sDK^*}$, $g_{D^*_sD^*K^*}$,
$g_{D_{s1}D_1K^*}$, $g_{D_sD^*K^*}$, and $g_{D_{s0}D_1K^*}$ form
factors in $SU_{f}(3)$ symmetry with $m_c=1.26~\rm GeV$ and
$\delta=0.7~\rm GeV$.}\label{T35}
\begin{ruledtabular}
\begin{tabular}{ccccccc}
$\mbox{Form factor}$&$A$&$B$&$\mbox{Form factor}$&$A$&$B$\\
\hline $g^{K^*}_{D^*_{s0}D^*_0K^*}(Q^2)$&6.96&7.57&$g^{K^*}_{D^*_{s1}D_1K^*}(Q^2)$&4.69&26.55\\
$g^{D^*_0}_{D^*_{s0}D^*_0K^*}(Q^2)$&6.16& 27.73&$g^{D_1}_{D^*_{s1}D_1K^*}(Q^2)$&3.62&22.20\\
$g^{K^*}_{D_sDK^*}(Q^2)$&2.06&2.08&$g^{K^*}_{D_{s}D^*K^*}(Q^2)$&3.43&7.88\\
$g^{D}_{D_sDK^*}(Q^2)$&2.62&33.57&$g^{D^*}_{D_{s}D^*K^*}(Q^2)$&3.08& 16.78&\\
$g^{K^*}_{D^*_sD^*K^*}(Q^2)$&5.16&220.80&$g^{K^*}_{D^*_{s0}D_1K^*}(Q^2)$&9.02&29.52\\
$g^{D^*}_{D^*_sD^*K^*}(Q^2)$&4.81&49.58&$g^{D_1}_{D^*_{s0}D_1K^*}(Q^2)$&7.06&19.32\\
\end{tabular}
\end{ruledtabular}
\end{table}

Also considering the $SU_{f}(3)$ symmetry, we obtain the values of
the coupling constants in $m_c=1.26~\rm GeV$ shown in Table
\ref{T36}.
\begin{table}[th]
\caption{The coupling constant of the vertices ${D^*_{s0}D^*_0K^*}$,
${D_sDK^*}$, ${D^*_sD^*K^*}$, ${D_{s1}D_1K^*}$,
${D_sD^*K^*}$, and ${D_{s0}D_1K^*}$,  in $SU_{f}(3)$ symmetry, in $\rm GeV^{-1}$. }\label{T36}
\begin{ruledtabular}
\begin{tabular}{ccccccccc}
$g$&$\mbox{off-shell charmed }$&$\mbox{ off-shell $K^*$}$&$g$&$\mbox{ off-shell charmed}$&$\mbox{ off-shell $K^*$}$\\
\hline
$g_{D^*_{s0}D^*_0K^*}$&$7.53\pm0.82$&$7.73\pm0.63$&$g_{D_{s1}D_1K^*}$&$4.71\pm0.32$&$4.83\pm0.52$\\
$g_{D_sDK^*}$&$2.91\pm0.37$&$3.02\pm0.45$&$g_{D_sD^*K^*}$&$3.91\pm0.32$&$3.79\pm0.41$\\
$g_{D^*_sD^*K^*}$&$5.22\pm0.71$&$5.18\pm0.66$&$g_{D^*_{s0}D_1K^*}$&$9.56\pm 1.12$&$9.27\pm0.92$\\
\end{tabular}
\end{ruledtabular}
\end{table}

It is possible to compare the coupling constant values of
$g_{D_sDK^*}$, $g_{D_sD^*K^*}$ and $g_{D^*_sD^*K^*}$  with
$g_{DD\rho}$, $g_{DD^*\rho}$ and $g_{D^*D^*\rho}$ respectively, in
the $SU_{f}(3)$ symmetry consideration. Table \ref{T37} shows that results are reasonably consistent to each other.
\begin{table}[th]
\caption{Values of the  coupling constant using the LCSR, and 3PSR.} \label{T37}
\begin{ruledtabular}
\begin{tabular}{cccc}
$g$ & Ours &3PSR \cite{MChiapparini,Rodrigues3,RDMatheus}& LCSR
\cite{Wang}
\\ \hline
$g_{D_sDK^*}$ &$2.95\pm0.44$& $3.42\pm0.44$& $2.62\pm0.66$\\
$g_{D^*_sD^*K^*}$ &$5.20\pm0.70$&$6.60\pm0.30$ & $--$\\
$g_{D_sD^*K^*}$ &$3.81\pm0.39$&$4.11\pm0.44$ & $3.56\pm0.60$
\end{tabular}
\end{ruledtabular}
\end{table}

In summary, taking into account the contributions of the
quark-quark, quark-gluon and gluon-gluon  condensate corrections,
the strong form factors  $g_{D^*_{s0}D^*_0K^*}$, $g_{D_sDK^*}$,
$g_{D^*_sD^*K^*}$, $g_{D_{s1}D_1K^*}$, $g_{D_sD^*K^*}$, and
$g_{D_{s0}D_1K^*}$ were estimated within the 3PSR without and with the
$SU_{f}(3)$ symmetry. For instance, the dependence of the strong
form factors $g_{D_sDK^*}$ on the transferred momentum square $Q^2$
were plotted. Also the coupling constants of these vertices were
evaluated.

\section*{Acknowledgments}
Partial support of the Isfahan university of technology research council is appreciated.

\clearpage
\appendix
\begin{center}
{\Large \textbf{Appendix--A}}
\end{center}
\setcounter{equation}{0} \renewcommand{\theequation}

In this appendix, the explicit expressions of the coefficients in the spectral
densities are given as:
\begin{eqnarray*}
 I_0(s,s',q^2) &=& \frac{1}{4\lambda^\frac{1}{2}(s,s',q^2)},\nonumber \\
\lambda(a,b,c) &=& a^2+ b^2+ c^2- 2ac- 2bc- 2ac ,\nonumber \\
\Delta&=&s'+m_s^2-m_c^2,\nonumber \\
\Delta'&=&s'+m_c^2-m_s^2,\nonumber \\
\Delta'' &=& s+m_s^2,\nonumber \\
u &=& s+s'-q^2,\nonumber \\
C_1 &=& \frac{1}{\lambda(s,s',q^2)}[2 s' \Delta'' -u \Delta],\nonumber \\
C_2 &=& \frac{1}{\lambda(s,s',q^2)}[2 s \Delta -u \Delta'' ],\nonumber \\
A &=& -\frac{1}{2\lambda(s,s',q^2)}[4ss'm_s^2-s\Delta^2-s'\Delta''^2-m_s^2 u^2+u\Delta\Delta''],\nonumber \\
E_1 &=& \frac{1}{2\lambda^2(s,s',q^2)}[ 8 ss'^2 m_s^2\Delta''
-2 s'm_s^2 u^2 \Delta''-4  s s' m_s^2 u \Delta  + m_s^2 u^3 \Delta -2 s'^2 \Delta''^3 \nonumber\\
&+& 3 s' u \Delta \Delta''^2  -2s s' \Delta^2 \Delta''
-u^2 \Delta^2 \Delta'' +s u \Delta^3],\nonumber\\
E_2 &=& \frac{1}{2\lambda^2(s,s',q^2)}[ 8 s^2 s' m_s^2 \Delta-2 s
m_s^2  u^2 \Delta''- 4  s s' m_s^2 u \Delta'' +m_s^2 u ^3\Delta''-2 s^2 \Delta ^3  \nonumber\\
&+& 3 s u  \Delta^2 \Delta''-2 s s' \Delta \Delta''^2 -u^2 \Delta
\Delta''^2 +  s' u \Delta''^3  ],
\end{eqnarray*}
also $A'=A_{|_{m_c\leftrightarrow m_s}}, C'_1={C_1}_{|_{m_c\leftrightarrow m_s}}, C'_2={C_2}_{|_{m_c\leftrightarrow m_s}}, E'_1={E_1}_{|_{m_c\leftrightarrow m_s}}$, and $E'_2={E_2}_{|_{m_c\leftrightarrow m_s}}$.

\clearpage
\appendix
\begin{center}
{\Large \textbf{Appendix--B}}
\end{center}
\setcounter{equation}{0} \renewcommand{\theequation}

In this appendix, the explicit expressions of the coefficients of
the quark and gluon condensate contributions of the strong form
factors in the Borel transform scheme for all the vertices are
presented.
\begin{eqnarray*}
C_{D^*_{s0}D^*_0K^*}^{D^*_0}&=&(3\,{\frac
{m_{{s}}{q}^{2}}{{M_{{1}}}^{2}}}-3\,{\frac {m_{{s}}{m_{{c}}}
^{2}}{{M_{{1}}}^{2}}}+{\frac
{{m_{{0}}}^{2}m_{{c}}}{{M_{{1}}}^{2}}}-3 \,{\frac
{m_{{s}}{m_{{c}}}^{2}}{{M_{{2}}}^{2}}}-6\,{\frac {m_{{c}}{m_{
{s}}}^{2}}{{M_{{2}}}^{2}}}-3\,{\frac
{m_{{c}}{q}^{2}{m_{{s}}}^{2}}{{M_
{{1}}}^{2}{M_{{2}}}^{2}}}+3\,{\frac
{{m_{{c}}}^{3}{m_{{s}}}^{2}}{{M_{{ 1}}}^{2}{M_{{2}}}^{2}}}
\\&-&2\,{\frac
{{m_{{0}}}^{2}{m_{{c}}}^{3}}{{M_{{1} }}^{2}{M_{{2}}}^{2}}}+2\,{\frac
{{m_{{0}}}^{2}m_{{c}}{q}^{2}}{{M_{{1}}
}^{2}{M_{{2}}}^{2}}}+3\,{\frac
{{m_{{c}}}^{3}{m_{{s}}}^{2}}{{M_{{2}}}^ {4}}}-\frac{3}{2}\,{\frac
{{m_{{0}}}^{2}{m_{{c}}}^{3}}{{M_{{2}}}^{4}}})\times
e^{-\frac{m_c^2}{M_2^2}},
\end{eqnarray*}
\begin{eqnarray*}
C_{D_{s}DK^*}^{D}&=&(3\,{\frac
{m_{{s}}{m_{{c}}}^{2}}{{M_{{1}}}^{2}}}-3\,{\frac {m_{{s}}{q}
^{2}}{{M_{{1}}}^{2}}}+3\,{\frac
{m_{{c}}{m_{{s}}}^{2}}{{M_{{1}}}^{2}}} -\frac{5}{2}\,{\frac
{{m_{{0}}}^{2}m_{{c}}}{{M_{{1}}}^{2}}}+3\,{\frac {m_{{s}}
{m_{{c}}}^{2}}{{M_{{2}}}^{2}}}-3\,{\frac
{{m_{{0}}}^{2}m_{{c}}}{{M_{{2 }}}^{2}}}-\frac{1}{2}\,{\frac
{{m_{{0}}}^{2}{m_{{c}}}^{3}}{{M_{{1}}}^{2}{M_{{2
}}}^{2}}}\\&+&\frac{1}{2}\,{\frac
{{m_{{0}}}^{2}m_{{c}}{q}^{2}}{{M_{{1}}}^{2}{M_{{
2}}}^{2}}}+3\,{\frac
{{m_{{c}}}^{3}{m_{{s}}}^{2}}{{M_{{2}}}^{4}}}-\frac{3}{2} \,{\frac
{{m_{{0}}}^{2}{m_{{c}}}^{3}}{{M_{{2}}}^{4}}})\times
e^{-\frac{m_c^2}{M_2^2}},
\end{eqnarray*}
\begin{eqnarray*}
C_{D^*_{s}D^*K^*}^{D^*}&=&(-6\,{\frac
{m_{{s}}{m_{{c}}}^{2}}{{M_{{2}}}^{2}}}-2\,{\frac {{m_{{0}}}
^{2}m_{{c}}}{{M_{{2}}}^{2}}}+6\,{\frac
{m_{{c}}{m_{{s}}}^{2}}{{M_{{2}} }^{2}}}+3\,{\frac
{m_{{s}}{q}^{2}}{{M_{{2}}}^{2}}}+3\,{\frac {m_{{c}}{
q}^{2}{m_{{s}}}^{2}}{{M_{{1}}}^{2}{M_{{2}}}^{2}}}-{\frac
{{m_{{0}}}^{2
}m_{{c}}{q}^{2}}{{M_{{1}}}^{2}{M_{{2}}}^{2}}}-3\,{\frac
{{m_{{c}}}^{3}
{m_{{s}}}^{2}}{{M_{{1}}}^{2}{M_{{2}}}^{2}}}\\&+&{\frac
{{m_{{0}}}^{2}{m_{{ c}}}^{3}}{{M_{{1}}}^{2}{M_{{2}}}^{2}}}-3\,{\frac
{{m_{{c}}}^{3}{m_{{s} }}^{2}}{{M_{{2}}}^{4}}}+\frac{3}{2}\,{\frac
{{m_{{0}}}^{2}{m_{{c}}}^{3}}{{M_{{ 2}}}^{4}}}+3\,m_{{s}})\times
e^{-\frac{m_c^2}{M_2^2}},
\end{eqnarray*}
\begin{eqnarray*}
C_{D^*_{s1}D_1K^*}^{D_1}&=& (2\,{\frac
{{m_{{0}}}^{2}m_{{c}}}{{M_{{2}}}^{2}}}-6\,{\frac {m_{{s}}{m_
{{c}}}^{2}}{{M_{{2}}}^{2}}}+3\,{\frac
{m_{{s}}{q}^{2}}{{M_{{2}}}^{2}}} -6\,{\frac
{m_{{c}}{m_{{s}}}^{2}}{{M_{{2}}}^{2}}}-{\frac {{m_{{0}}}^{2
}{m_{{c}}}^{3}}{{M_{{1}}}^{2}{M_{{2}}}^{2}}}-3\,{\frac
{m_{{c}}{q}^{2} {m_{{s}}}^{2}}{{M_{{1}}}^{2}{M_{{2}}}^{2}}}+{\frac
{{m_{{0}}}^{2}m_{{c
}}{q}^{2}}{{M_{{1}}}^{2}{M_{{2}}}^{2}}}\\&+&3\,{\frac
{{m_{{c}}}^{3}{m_{{s }}}^{2}}{{M_{{1}}}^{2}{M_{{2}}}^{2}}}+3\,{\frac
{{m_{{c}}}^{3}{m_{{s}} }^{2}}{{M_{{2}}}^{4}}}-\frac{3}{2}\,{\frac
{{m_{{0}}}^{2}{m_{{c}}}^{3}}{{M_{{2 }}}^{4}}}+3\,m_{{s}})\times
e^{-\frac{m_c^2}{M_2^2}},
\end{eqnarray*}
\begin{eqnarray*}
C_{D_{s}D^* K^*}^{D^*}&=&( 6\,{\frac
{m_{{s}}m_{{c}}}{{M_{{2}}}^{2}}}-6\,{\frac {{q}^{2}{m_{{s}}
}^{2}}{{M_{{1}}}^{2}{M_{{2}}}^{2}}}+2\,{\frac
{{m_{{0}}}^{2}{q}^{2}}{{ M_{{1}}}^{2}{M_{{2}}}^{2}}}+6\,{\frac
{{m_{{c}}}^{2}{m_{{s}}}^{2}}{{M_ {{1}}}^{2}{M_{{2}}}^{2}}}-2\,{\frac
{{m_{{0}}}^{2}{m_{{c}}}^{2}}{{M_{{ 1}}}^{2}{M_{{2}}}^{2}}}+6\,{\frac
{{m_{{c}}}^{2}{m_{{s}}}^{2}}{{M_{{2} }}^{4}}}\\&-&3\,{\frac
{{m_{{0}}}^{2}{m_{{c}}}^{2}}{{M_{{2}}}^{4}}})\times
e^{-\frac{m_c^2}{M_2^2}},
\end{eqnarray*}
\begin{eqnarray*}
C_{D^*_{s0}D_1K^*}^{D_1}&=&(-6\,{\frac
{m_{{s}}m_{{c}}}{{M_{{2}}}^{2}}}-6\,{\frac {{q}^{2}{m_{{s}}
}^{2}}{{M_{{1}}}^{2}{M_{{2}}}^{2}}}+2\,{\frac
{{m_{{0}}}^{2}{q}^{2}}{{ M_{{1}}}^{2}{M_{{2}}}^{2}}}+6\,{\frac
{{m_{{c}}}^{2}{m_{{s}}}^{2}}{{M_ {{1}}}^{2}{M_{{2}}}^{2}}}-2\,{\frac
{{m_{{0}}}^{2}{m_{{c}}}^{2}}{{M_{{ 1}}}^{2}{M_{{2}}}^{2}}}+6\,{\frac
{{m_{{c}}}^{2}{m_{{s}}}^{2}}{{M_{{2} }}^{4}}}\\&-&3\,{\frac
{{m_{{0}}}^{2}{m_{{c}}}^{2}}{{M_{{2}}}^{4}}})\times
e^{-\frac{m_c^2}{M_2^2}},
\end{eqnarray*}

\begin{eqnarray*}
C_{D^*_{s0}D^*_{0}K^*}^{K^*}&=&\hat{I}_{2}(3,2,2)m_{c}^{6}+\hat{I}_{0}(3,2,2)m_{c}^{6}-\hat{I}_{1}(3,2,2)m_{c}^{6}+\hat{I}_{2}(3,2,2)m_{c}^{5}m_{s}
\\
&&+2\hat{I}_{0}(3,2,2)m_{c}^{5}m_{s}-\hat{I}_{1}(3,2,2)m_{c}^{5}m_{s}-\hat{I}_{2}(3,2,2)m_{c}^{4}m_{s}^{2}+\hat{I}_{1}(3,2,2)m_{c}^{4}m_{s}^{2}
\\
&&+\hat{I}_{1}(3,2,2)m_{c}^{3}m_{s}^{3}-\hat{I}_{2}(3,2,2)m_{c}^{3}m_{s}^{3}-\hat{I}_{0}(3,2,2)m_{c}^{2}m_{s}^{4}+3\hat{I}_{0}(2,2,2)m_{c}^{4}
\\
&&+\hat{I}_{2}(3,2,1)m_{c}^{4}+\hat{I}_0^{[1,0]}(3,2,2)m_{c}^{4}-\hat{I}_{1}(3,2,1)m_{c}^{4}+6\hat{I}_{0}(4,1,1)m_{c}^{3}m_{s}
\\
&&-\hat{I}_{1}(3,1,2)m_{c}^{3}m_{s}+2\hat{I}_{2}(2,2,2)m_{c}^{3}m_{s}+\hat{I}_{2}(3,1,2)m_{c}^{3}m_{s}+2\hat{I}_0^{[1,0]}(3,2,2)m_{c}^{3}m_{s}
\\
&&+2\hat{I}_{0}(3,2,1)m_{c}^{3}m_{s}-2\hat{I}_{1}(2,2,2)m_{c}^{3}m_{s}+8\hat{I}_{0}(3,1,2)m_{c}^{2}m_{s}^{2}-\hat{I}_1^{[1,0]}(3,2,2)m_{c}^{2}m_{s}^{2}
\\
&&+6\hat{I}_{1}(1,1,4)m_{c}^{2}m_{s}^{2}+3\hat{I}_{0}(4,1,1)m_{c}^{2}m_{s}^{2}-6\hat{I}_{2}(1,1,4)m_{c}^{2}m_{s}^{2}-6\hat{I}_{0}(1,1,4)m_{c}^{2}m_{s}^{2}
\\
&&-\hat{I}_2^{[1,0]}(3,2,2)m_{c}^{2}m_{s}^{2}-\hat{I}_2^{[1,0]}(3,2,2)m_{c}m_{s}^{3}+2\hat{I}_{2}(3,1,2)m_{c}m_{s}^{3}-\hat{I}_1^{[1,0]}(3,2,2)m_{c}m_{s}^{3}
\\
&&-12\hat{I}_{0}(1,1,4)m_{c}m_{s}^{3}-2\hat{I}_{1}(3,1,2)m_{c}m_{s}^{3}-2\hat{I}_0^{[1,0]}(3,2,2)m_{c}m_{s}^{3}+8\hat{I}_{0}(3,1,2)m_{c}m_{s}^{3}
\\
&&-6\hat{I}_{0}(1,1,4)m_{s}^{4}+3\hat{I}_{0}(3,1,2)m_{s}^{4}+3\hat{I}_{0}(2,2,1)m_{c}^{2}-3\hat{I}_1^{[0,1]}(3,1,2)m_{c}^{2}
\\
&&-2\hat{I}_{1}(1,2,2)m_{c}^{2}-3\hat{I}_0^{[0,1]}(4,1,1)m_{c}^{2}+\hat{I}_0^{[0,1]}(3,2,1)m_{c}^{2}-3\hat{I}_2^{[0,1]}(3,1,2)m_{c}^{2}
\\
&&-\hat{I}_0^{[0,2]}(3,2,2)m_{c}^{2}+5\hat{I}_1^{[1,1]}(3,2,2)m_{c}^{2}+5\hat{I}_2^{[1,1]}(3,2,2)m_{c}^{2}+2\hat{I}_{2}(1,2,2)m_{c}^{2}
\\
&&+2\hat{I}_{1}(1,2,2)m_{c}m_{s}-3\hat{I}_2^{[1,0]}(3,2,1)m_{c}m_{s}-2\hat{I}_{2}(1,2,2)m_{c}m_{s}-\hat{I}_{2}(2,2,1)m_{c}m_{s}
\\
&&+\hat{I}_1^{[1,0]}(3,2,2)m_{c}m_{s}+2\hat{I}_0^{[1,1]}(3,2,2)m_{c}m_{s}+6\hat{I}_{0}(3,1,1)m_{c}m_{s}+\hat{I}_{1}(2,2,1)m_{c}m_{s}
\\
&&-2\hat{I}_1^{[0,1]}(2,1,3)m_{c}m_{s}-2\hat{I}_2^{[0,1]}(2,1,3)m_{c}m_{s}+\hat{I}_2^{[1,1]}(3,2,2)m_{c}m_{s}-3\hat{I}_1^{[1,0]}(3,2,1)m_{c}m_{s}
\\
&&-\hat{I}_1^{[1,0]}(2,2,2)m_{s}^{2}-3\hat{I}_0^{[1,0]}(3,2,1)m_{s}^{2}+6\hat{I}_0^{[0,1]}(1,1,4)m_{s}^{2}+4\hat{I}_{0}(3,1,1)m_{s}^{2}
\\
&&-5\hat{I}_{0}(2,1,2)m_{s}^{2}-\hat{I}_2^{[1,0]}(2,2,2)m_{s}^{2}-2\hat{I}_0^{[1,0]}(2,2,2)m_{s}^{2}-\hat{I}_{2}(2,1,2)m_{s}^{2}+\hat{I}_{1}(2,1,2)m_{s}^{2}
\\
&&-6\hat{I}_0^{[0,1]}(3,1,2)m_{s}^{2}-\hat{I}_0^{[1,0]}(1,2,2)+2\hat{I}_0^{[1,1]}(2,2,2)-2\hat{I}_0^{[0,1]}(2,2,1) \\
&&-2\hat{I}_{2}(2,1,1)+3\hat{I}_{0}(2,1,1)-\hat{I}_0^{[0,1]}(1,2,2)-6\hat{I}_0^{[0,1]}(1,3,1)+2\hat{I}_{1}(2,1,1)
\\
&&+2\hat{I}_{1}(1,2,1)+2\hat{I}_{1}(1,1,2)-\hat{I}_{0}(1,2,1)-\hat{I}_0^{[1,0]}(2,2,1)-2\hat{I}_{2}(1,2,1) \\
&&+\hat{I}_0^{[0,2]}(2,2,2)-2\hat{I}_{2}(1,1,2)-3\hat{I}_0^{[0,1]}(2,1,2),
\end{eqnarray*}
\begin{eqnarray*}
C_{D_sDK^*}^{K^*} &=&-\hat{I}_0(3,2,2)m_c^6-\hat{I}_1(3,2,2)m_c^5 m_s+\hat{I}_2(3,2,2)m_c^5m_s+2\hat{I}_0(3,2,2)m_c^5m_s \\
&&-\hat{I}_1(3,2,2)m_c^4m_s^2+\hat{I}_2(3,2,2)m_c^4m_s^2-2\hat{I}_0(3,2,2)m_c^3m_s^3-\hat{I}_2(3,1,2)m_c^4\\
&&-\hat{I}_0^{[0,1]}(3,2,2)m_c^4+3\hat{I}_1(2,2,2)m_c^4-3\hat{I}_2(2,2,2)m_c^4+\hat{I}_1(3,1,2)m_c^4 \\
&&-3\hat{I}_0(3,2,1)m_c^4+2\hat{I}_2(2,2,2)m_c^3m_s+2\hat{I}_0(3,2,1)m_c^3m_s+2\hat{I}_0^{[1,0]}(3,2,2)m_c^3m_s \\
&&-2\hat{I}_1(2,2,2)m_c^3m_s+\hat{I}_2^{[1,0]}(3,2,2)m_c^3m_s+6\hat{I}_0(4,1,1)m_c^3m_s+4\hat{I}_0(2,2,2)m_c^3m_s \\
&&+\hat{I}_2(3,2,1)m_c^3m_s-\hat{I}_1(3,2,1)m_c^3m_s+\hat{I}_0(3,1,2)m_c^3m_s+\hat{I}_1^{[1,0]}(3,2,2)m_c^3m_s \\
&&-2\hat{I}_0(2,2,2)m_c^2m_s^2+2\hat{I}_1(3,1,2)m_c^2m_s^2-3\hat{I}_0(4,1,1)m_c^2m_s^2-2\hat{I}_2(3,1,2)m_c^2m_s^2 \\
&&-2\hat{I}_0^{[0,1]}(3,2,2)m_c^2m_s^2+5\hat{I}_0(3,1,2)m_cm_s^3+2\hat{I}_2(2,1,3)m_cm_s^3+2\hat{I}_2(3,1,2)m_cm_s^3 \\
&&-2\hat{I}_1(2,1,3)m_cm_s^3-2\hat{I}_1(3,1,2)m_cm_s^3-\hat{I}_0(2,2,2)m_s^4+\hat{I}_0^{[1,0]}(3,2,2)m_s^4 \\
&&-5\hat{I}_1^{[1,1]}(3,2,2)m_c^2+2\hat{I}_0^{[0,1]}(3,1,2)m_c^2-5\hat{I}_2^{[1,1]}(3,2,2)m_c^2+3\hat{I}_2^{[1,0]}(3,2,1)m_c^2 \\
&&+2\hat{I}_0^{[0,1]}(2,2,2)m_c^2+3\hat{I}_1(1,3,1)m_c^2-2\hat{I}_0(1,2,2)m_c^2+3\hat{I}_1^{[1,0]}(3,2,1)m_c^2 \\
&&-3\hat{I}_2(1,3,1)m_c^2-12\hat{I}_0(1,1,3)m_cm_s+\hat{I}_2^{[1,1]}(3,2,2)m_cm_s-3\hat{I}_1^{[1,0]}(3,2,1)m_cm_s \\
&&-9\hat{I}_0(2,1,2)m_cm_s+3\hat{I}_1(1,3,1)m_cm_s-2\hat{I}_1^{[0,1]}(2,1,3)m_cm_s-\hat{I}_2(2,2,1)m_cm_s \\
&&-3\hat{I}_2^{[1,0]}(3,2,1)m_cm_s+\hat{I}_1(2,2,1)m_cm_s+4\hat{I}_1(2,1,2)m_cm_s-2\hat{I}_2^{[0,1]}(2,1,3)m_cm_s \\
&&-4\hat{I}_2(2,1,2)m_cm_s-3\hat{I}_2(1,3,1)m_cm_s+\hat{I}_1^{[1,1]}(3,2,2)m_cm_s-2\hat{I}_0(3,1,1)m_s^2 \\
&&+\hat{I}_2(2,1,2)m_s^2-\hat{I}_1(2,1,2)m_s^2+2\hat{I}_0^{[1,0]}(2,2,2)m_s^2+3\hat{I}_0(1,1,3)m_s^2 \\
&&+\hat{I}_2^{[1,0]}(2,2,2)m_s^2-2\hat{I}_0(2,2,1)m_s^2+\hat{I}_1^{[1,0]}(2,2,2)m_s^2+2\hat{I}_0^{[0,1]}(2,2,2)m_s^2 \\
&&+3\hat{I}_0(2,1,2)m_s^2-2\hat{I}_0^{[1,1]}(2,2,2)-2\hat{I}_0^{[0,2]}(3,1,2)-2\hat{I}_1(1,2,1)+2\hat{I}_0^{[0,1]}(2,2,1) \\
&&+\hat{I}_0^{[0,1]}(1,2,2)+2\hat{I}_2(1,2,1)-3\hat{I}_0^{[1,1]}(3,2,1)-3\hat{I}_0^{[0,1]}(1,1,3)+\hat{I}_0(2,1,1) \\
&&+2\hat{I}_2(1,1,2)+\hat{I}_0^{[0,1]}(2,1,2)-2\hat{I}_1(1,1,2),
\end{eqnarray*}
\begin{eqnarray*}
C_{D^*_{s}D^*K^*}^{K^*}&=&\hat{I}_{0}(3,2,2)m_{c}^{6}-\hat{I}_{0}(3,2,2)m_{c}^{5}m_{s}+\hat{I}_{0}(3,2,2)m_{c}^{3}m_{s}^{3}+\hat{I}_1^{[1,0]}(3,2,2)m_{c}^{4}
\\
&&+3\hat{I}_{0}(4,1,1)m_{c}^{4}-\hat{I}_2^{[1,0]}(3,2,2)m_{c}^{4}+2\hat{I}_{6}(3,2,2)m_{c}^{4}+2\hat{I}_{2}(2,1,3)m_{c}^{3}m_{s}
\\
&&+2\hat{I}_{0}(2,1,3)m_{c}^{3}m_{s}-2\hat{I}_{1}(2,2,2)m_{c}^{3}m_{s}-2\hat{I}_{1}(2,1,3)m_{c}^{3}m_{s}-\hat{I}_{1}(3,2,1)m_{c}^{3}m_{s}
\\
&&+\hat{I}_{2}(3,2,1)m_{c}^{3}m_{s}-3\hat{I}_{0}(4,1,1)m_{c}^{3}m_{s}+2\hat{I}_{2}(2,2,2)m_{c}^{3}m_{s}+\hat{I}_{0}(2,2,2)m_{c}^{2}m_{s}^{2}
\\
&&-2\hat{I}_{6}(3,2,2)m_{c}^{2}m_{s}^{2}+\hat{I}_0^{[0,1]}(3,2,2)m_{c}^{2}m_{s}^{2}+\hat{I}_2^{[1,0]}(3,2,2)m_{c}^{2}m_{s}^{2}-\hat{I}_1^{[1,0]}(3,2,2)m_{c}^{2}m_{s}^{2}
\\
&&+2\hat{I}_{2}(3,1,2)m_{c}m_{s}^{3}-\hat{I}_2^{[1,0]}(3,2,2)m_{c}m_{s}^{3}-2\hat{I}_{1}(3,1,2)m_{c}m_{s}^{3}+6\hat{I}_{0}(1,1,4)m_{c}m_{s}^{3}
\\
&&+\hat{I}_1^{[1,0]}(3,2,2)m_{c}m_{s}^{3}+\hat{I}_{0}(3,1,2)m_{s}^{4}+2\hat{I}_6^{[1,0]}(3,2,2)m_{c}^{2}+\hat{I}_{0}(2,1,2)m_{c}^{2}
\\
&&+8\hat{I}_{8}(3,2,1)m_{c}^{2}+4\hat{I}_8^{[0,1]}(3,2,2)m_{c}^{2}+2\hat{I}_6^{[0,1]}(3,2,2)m_{c}^{2}-4\hat{I}_{6}(3,2,1)m_{c}^{2}
\\
&&-4\hat{I}_7^{[0,1]}(3,2,2)m_{c}^{2}+3\hat{I}_1^{[1,0]}(4,1,1)m_{c}^{2}+6\hat{I}_{6}(4,1,1)m_{c}^{2}+6\hat{I}_{6}(3,1,2)m_{c}^{2}
\\
&&-8\hat{I}_{7}(3,2,1)m_{c}^{2}-12\hat{I}_{7}(4,1,1)m_{c}^{2}+2\hat{I}_{0}(2,2,1)m_{c}^{2}+12\hat{I}_{8}(4,1,1)m_{c}^{2}
\\
&&-3\hat{I}_2^{[1,0]}(4,1,1)m_{c}^{2}+8\hat{I}_{6}(2,1,3)m_{c}m_{s}+4\hat{I}_{6}(2,2,2)m_{c}m_{s}-\hat{I}_0^{[1,1]}(3,2,2)m_{c}m_{s}
\\
&&-3\hat{I}_{1}(1,3,1)m_{c}m_{s}-2\hat{I}_1^{[0,1]}(3,1,2)m_{c}m_{s}+3\hat{I}_1^{[1,0]}(3,2,1)m_{c}m_{s}-2\hat{I}_{6}(3,2,1)m_{c}m_{s}
\\
&&+3\hat{I}_{2}(1,3,1)m_{c}m_{s}+2\hat{I}_2^{[0,1]}(3,1,2)m_{c}m_{s}-\hat{I}_{0}(2,1,2)m_{c}m_{s}-2\hat{I}_{6}(3,1,2)m_{c}m_{s}
\\
&&-3\hat{I}_2^{[1,0]}(3,2,1)m_{c}m_{s}-2\hat{I}_6^{[1,0]}(3,2,2)m_{s}^{2}-12\hat{I}_{6}(1,1,4)m_{s}^{2}-\hat{I}_{0}(2,1,2)m_{s}^{2}
\\
&&-3\hat{I}_{0}(1,1,3)m_{s}^{2}-2\hat{I}_0^{[0,1]}(3,1,2)m_{s}^{2}+3\hat{I}_0^{[1,0]}(1,1,4)m_{s}^{2}+\hat{I}_0^{[1,1]}(3,2,2)m_{s}^{2}
\\
&&+6\hat{I}_2^{[1,0]}(1,1,4)m_{s}^{2}+\hat{I}_{0}(3,1,1)m_{s}^{2}+24\hat{I}_{7}(1,1,4)m_{s}^{2}-24\hat{I}_{8}(1,1,4)m_{s}^{2}
\\
&&+4\hat{I}_{6}(2,2,2)m_{s}^{2}-6\hat{I}_1^{[1,0]}(1,1,4)m_{s}^{2}-2\hat{I}_0^{[1,0]}(2,2,2)m_{s}^{2}+3\hat{I}_0^{[0,1]}(1,1,4)m_{s}^{2}
\\
&&+2\hat{I}_{6}(3,1,2)m_{s}^{2}+3\hat{I}_1^{[0,1]}(2,1,2)-2\hat{I}_{2}(2,1,1)+3\hat{I}_{6}(1,3,1) \\
&&-3\hat{I}_2^{[0,1]}(2,1,2)-4S_{1},1(3,2,2)+4\hat{I}_7^{[1,0]}(3,2,1)-8\hat{I}_{8}(2,2,1) \\
&&-2\hat{I}_6^{[0,1]}(2,2,2)-2\hat{I}_6^{[0,1]}(3,1,2)+8\hat{I}_{7}(2,2,1)-4\hat{I}_8^{[1,0]}(3,2,1) \\
&&-4\hat{I}_6^{[1,0]}(3,2,1)+2\hat{I}_6^{[1,1]}(3,2,2)-12\hat{I}_{8}(3,1,1)+3\hat{I}_1^{[1,0]}(3,1,2) \\
&&+2\hat{I}_{1}(2,1,1)-2\hat{I}_2^{[1,0]}(1,2,2)+2\hat{I}_1^{[1,0]}(1,2,2)+4\hat{I}_7^{[0,1]}(2,2,2) \\
&&-4\hat{I}_8^{[0,1]}(2,2,2)+4\hat{I}_{6}(2,2,1)-\hat{I}_0^{[0,1]}(3,1,1)-3\hat{I}_2^{[1,1]}(3,1,2) \\
&&+\hat{I}_2^{[2,0]}(2,2,2)-\hat{I}_1^{[2,0]}(2,2,2)+12\hat{I}_{7}(3,1,1)-8\hat{I}_{6}(2,1,2) \\
&&+4\hat{I}_8^{[1,1]}(3,2,2)+\hat{I}_0^{[2,0]}(3,2,1)-2\hat{I}_0^{[1,0]}(1,2,2)-\hat{I}_0^{[0,1]}(2,1,2) \\
&&-3\hat{I}_{0}(2,1,1)-4\hat{I}_{6}(3,1,1)-2\hat{I}_6^{[1,0]}(2,2,2)+4\hat{I}_7^{[1,0]}(2,2,2) \\
&&-4\hat{I}_8^{[1,0]}(2,2,2)+2\hat{I}_{0}(1,1,2)-\hat{I}_0^{[1,0]}(3,1,1)-6\hat{I}_{7}(1,3,1)\\
&&+6\hat{I}_{8}(1,3,1),
\end{eqnarray*}
\begin{eqnarray*}
C_{D_{s1}D_1K^*}^{K^*}&=&\hat{I}_{0}(3,2,2)m_{c}^{6}+\hat{I}_{2}(3,2,2)m_{c}^{3}m_{s}^{3}-\hat{I}_{1}(3,2,2)m_{c}^{3}m_{s}^{3}-\hat{I}_{0}(3,2,2)m_{c}^{3}m_{s}^{3}
\\
&&-4\hat{I}_{7}(3,2,2)m_{c}^{4}+3\hat{I}_{0}(2,2,2)m_{c}^{4}+2\hat{I}_{6}(3,2,2)m_{c}^{4}+4\hat{I}_{8}(3,2,2)m_{c}^{4}
\\
&&+3\hat{I}_{0}(4,1,1)m_{c}^{4}-2\hat{I}_{2}(2,2,2)m_{c}^{3}m_{s}+2\hat{I}_{1}(2,2,2)m_{c}^{3}m_{s}-2\hat{I}_{2}(2,1,3)m_{c}^{3}m_{s}
\\
&&+2\hat{I}_{1}(2,1,3)m_{c}^{3}m_{s}+\hat{I}_{1}(3,2,1)m_{c}^{3}m_{s}+\hat{I}_0^{[0,1]}(3,2,2)m_{c}^{3}m_{s}-\hat{I}_{2}(3,2,1)m_{c}^{3}m_{s}
\\
&&-\hat{I}_{0}(3,1,2)m_{c}^{3}m_{s}-2\hat{I}_{6}(3,2,2)m_{c}^{2}m_{s}^{2}-\hat{I}_1^{[1,0]}(3,2,2)m_{c}^{2}m_{s}^{2}-6\hat{I}_{0}(1,1,4)m_{c}^{2}m_{s}^{2}
\\
&&+\hat{I}_2^{[1,0]}(3,2,2)m_{c}^{2}m_{s}^{2}+6\hat{I}_{2}(1,1,4)m_{c}m_{s}^{3}-6\hat{I}_{1}(1,1,4)m_{c}m_{s}^{3}+\hat{I}_{0}(3,1,2)m_{s}^{4}
\\
&&+\hat{I}_{1}(2,1,2)m_{c}^{2}-\hat{I}_{1}(3,1,1)m_{c}^{2}+2\hat{I}_{0}(2,2,1)m_{c}^{2}+\hat{I}_1^{[2,0]}(3,2,2)m_{c}^{2}
\\
&&+8\hat{I}_{8}(3,2,1)m_{c}^{2}+4\hat{I}_8^{[0,1]}(3,2,2)m_{c}^{2}+2\hat{I}_6^{[0,1]}(3,2,2)m_{c}^{2}-12\hat{I}_{7}(4,1,1)m_{c}^{2}
\\
&&-4\hat{I}_7^{[0,1]}(3,2,2)m_{c}^{2}+4\hat{I}_8^{[1,0]}(3,2,2)m_{c}^{2}+\hat{I}_{2}(3,1,1)m_{c}^{2}+2\hat{I}_8^{[1,0]}(3,2,2)m_{c}^{2}
\\
&&-\hat{I}_2^{[2,0]}(3,2,2)m_{c}^{2}-4\hat{I}_{6}(3,2,1)m_{c}^{2}-8\hat{I}_{7}(3,2,1)m_{c}^{2}-\hat{I}_0^{[0,1]}(2,2,2)m_{c}^{2}
\\
&&+\hat{I}_{0}(2,1,2)m_{c}^{2}+6\hat{I}_{6}(4,1,1)m_{c}^{2}-\hat{I}_{2}(2,1,2)m_{c}^{2}-4\hat{I}_7^{[1,0]}(3,2,2)m_{c}^{2}
\\
&&+6\hat{I}_{6}(3,1,2)m_{c}^{2}+12\hat{I}_{8}(4,1,1)m_{c}^{2}-4\hat{I}_{1}(1,1,3)m_{c}m_{s}+4\hat{I}_{2}(1,1,3)m_{c}m_{s}
\\
&&-4\hat{I}_{6}(2,2,2)m_{c}m_{s}-3\hat{I}_{2}(1,3,1)m_{c}m_{s}+2\hat{I}_{0}(1,2,2)m_{c}m_{s}+2\hat{I}_{6}(3,1,2)m_{c}m_{s}
\\
&&+\hat{I}_{0}(2,1,2)m_{c}m_{s}-2\hat{I}_2^{[0,1]}(3,1,2)m_{c}m_{s}-8\hat{I}_{6}(2,1,3)m_{c}m_{s}+2\hat{I}_0^{[1,0]}(2,1,3)m_{c}m_{s}
\\
&&+2\hat{I}_{6}(3,2,1)m_{c}m_{s}+\hat{I}_{2}(2,1,2)m_{c}m_{s}+2\hat{I}_1^{[0,1]}(3,1,2)m_{c}m_{s}-\hat{I}_{1}(2,1,2)m_{c}m_{s}
\\
&&+3\hat{I}_{1}(1,3,1)m_{c}m_{s}+3\hat{I}_{0}(2,2,1)m_{c}m_{s}-3\hat{I}_{0}(1,1,3)m_{s}^{2}+4\hat{I}_{6}(2,2,2)m_{s}^{2}
\\
&&-6\hat{I}_1^{[1,0]}(1,1,4)m_{s}^{2}+3\hat{I}_0^{[0,1]}(1,1,4)m_{s}^{2}-12\hat{I}_{6}(1,1,4)m_{s}^{2}-2\hat{I}_0^{[0,1]}(3,1,2)m_{s}^{2}
\\
&&-\hat{I}_{0}(2,1,2)m_{s}^{2}+6\hat{I}_2^{[1,0]}(1,1,4)m_{s}^{2}+2\hat{I}_{6}(3,1,2)m_{s}^{2}-2\hat{I}_6^{[1,0]}(3,2,2)m_{s}^{2}
\\
&&-2\hat{I}_2^{[1,0]}(1,2,2)+\hat{I}_2^{[1,0]}(3,1,1)-4\hat{I}_6^{[1,0]}(3,2,1)-\hat{I}_0^{[1,0]}(3,1,1) \\
&&-8\hat{I}_{8}(2,2,1)-\hat{I}_0^{[0,1]}(2,1,2)-12\hat{I}_{8}(3,1,1)-2\hat{I}_6^{[1,0]}(2,2,2)+12\hat{I}_{7}(3,1,1)
\\
&&-8\hat{I}_{8}(2,1,2)-8\hat{I}_{6}(2,1,2)-\hat{I}_1^{[1,0]}(3,1,1)+8\hat{I}_{7}(2,1,2)+4\hat{I}_7^{[1,0]}(3,2,1)
\\
&&-4\hat{I}_8^{[1,0]}(3,2,1)+4\hat{I}_7^{[0,1]}(3,1,2)-4\hat{I}_8^{[0,1]}(3,1,2)+2\hat{I}_1^{[0,1]}(1,2,2) \\
&&+2\hat{I}_{1}(2,1,1)-4\hat{I}_7^{[1,1]}(3,2,2)+4\hat{I}_8^{[1,1]}(3,2,2)+3\hat{I}_{6}(1,3,1)-2\hat{I}_6^{[0,1]}(3,1,2)
\\
&&-2\hat{I}_6^{[0,1]}(2,2,2)-4\hat{I}_{6}(3,1,1)+\hat{I}_0^{[0,2]}(3,1,2)-\hat{I}_0^{[0,1]}(3,1,1) \\
&&-3\hat{I}_{0}(2,1,1)+4\hat{I}_{6}(2,2,1)+2\hat{I}_6^{[1,1]}(3,2,2)-2\hat{I}_{2}(2,1,1) \\
&&+8\hat{I}_{7}(2,2,1)-2\hat{I}_0^{[0,1]}(2,2,1),
\end{eqnarray*}
\begin{eqnarray*}
C_{D_{s}D^*K^*}^{K^*}&=&2\hat{I}_{2}(3,2,2)m_{c}^{5}+2\hat{I}_{0}(3,2,2)m_{c}^{5}+2\hat{I}_{1}(3,2,2)m_{c}^{5}-2\hat{I}_{2}(3,2,2)m_{c}^{4}m_{s}
\\
&&-2\hat{I}_{0}(3,2,2)m_{c}^{3}m_{s}^{2}-2\hat{I}_{2}(3,2,2)m_{c}^{3}m_{s}^{2}-2\hat{I}_{1}(3,2,2)m_{c}^{3}m_{s}^{2}+2\hat{I}_{2}(3,2,2)m_{c}^{2}m_{s}^{3}
\\
&&+4\hat{I}_{2}(2,2,2)m_{c}^{3}+2\hat{I}_0^{[0,1]}(3,2,2)m_{c}^{3}+6\hat{I}_{1}(4,1,1)m_{c}^{3}+2\hat{I}_2^{[1,0]}(3,2,2)m_{c}^{3}
\\
&&+6\hat{I}_{0}(4,1,1)m_{c}^{3}+4\hat{I}_{0}(2,2,2)m_{c}^{3}+2\hat{I}_{0}(3,2,1)m_{c}^{3}+2\hat{I}_0^{[1,0]}(3,2,2)m_{c}^{3}
\\
&&+2\hat{I}_1^{[0,1]}(3,2,2)m_{c}^{3}+2\hat{I}_2^{[0,1]}(3,2,2)m_{c}^{3}-2\hat{I}_{1}(3,1,2)m_{c}^{3}+2\hat{I}_{1}(3,2,1)m_{c}^{3}
\\
&&+2\hat{I}_1^{[1,0]}(3,2,2)m_{c}^{3}+4\hat{I}_{1}(2,2,2)m_{c}^{3}+6\hat{I}_{2}(4,1,1)m_{c}^{3}-4\hat{I}_{2}(2,2,2)m_{c}^{2}m_{s}
\\
&&-2\hat{I}_2^{[1,0]}(3,2,2)m_{c}^{2}m_{s}+8\hat{I}_{0}(2,1,3)m_{c}^{2}m_{s}+2\hat{I}_{1}(3,1,2)m_{c}^{2}m_{s}-2\hat{I}_{0}(3,1,2)m_{c}^{2}m_{s}
\\
&&+4\hat{I}_{2}(2,1,3)m_{c}^{2}m_{s}-2\hat{I}_{2}(3,1,2)m_{c}^{2}m_{s}-6\hat{I}_{2}(4,1,1)m_{c}^{2}m_{s}-2\hat{I}_2^{[0,1]}(3,2,2)m_{c}^{2}m_{s}
\\
&&+4\hat{I}_{1}(3,1,2)m_{c}m_{s}^{2}+4\hat{I}_{0}(3,1,2)m_{c}m_{s}^{2}-2\hat{I}_1^{[1,0]}(3,2,2)m_{c}m_{s}^{2}+12\hat{I}_{1}(1,1,4)m_{c}m_{s}^{2}
\\
&&-2\hat{I}_2^{[1,0]}(3,2,2)m_{c}m_{s}^{2}+12\hat{I}_{2}(1,1,4)m_{c}m_{s}^{2}-2\hat{I}_0^{[1,0]}(3,2,2)m_{c}m_{s}^{2}+6\hat{I}_{2}(3,1,2)m_{c}m_{s}^{2}
\\
&&+12\hat{I}_{0}(1,1,4)m_{c}m_{s}^{2}-4\hat{I}_{2}(3,1,2)m_{s}^{3}+2\hat{I}_2^{[1,0]}(3,2,2)m_{s}^{3}-12\hat{I}_{2}(1,1,4)m_{s}^{3}
\\
&&-4\hat{I}_{2}(2,1,3)m_{s}^{3}-2\hat{I}_{2}(2,2,2)m_{s}^{3}+2\hat{I}_{2}(3,1,1)m_{c}-4\hat{I}_2^{[1,0]}(3,2,1)m_{c}
\\
&&-4\hat{I}_1^{[0,1]}(3,1,2)m_{c}+4\hat{I}_{0}(1,2,2)m_{c}-2\hat{I}_1^{[1,0]}(3,1,2)m_{c}+4\hat{I}_{2}(1,2,2)m_{c}
\\
&&+8\hat{I}_{0}(2,1,2)m_{c}+8\hat{I}_{2}(2,1,2)m_{c}+2\hat{I}_{0}(3,1,1)m_{c}+4\hat{I}_{1}(1,2,2)m_{c}
\\
&&-6\hat{I}_1^{[1,0]}(3,2,1)m_{c}+2\hat{I}_{0}(2,2,1)m_{c}-4\hat{I}_0^{[0,1]}(3,1,2)m_{c}-4\hat{I}_2^{[0,1]}(3,1,2)m_{c}
\\
&&+2\hat{I}_2^{[1,1]}(3,2,2)m_{c}+2\hat{I}_1^{[1,1]}(3,2,2)m_{c}+2\hat{I}_{1}(2,2,1)m_{c}+6\hat{I}_{1}(2,1,2)m_{c}
\\
&&+2\hat{I}_0^{[1,1]}(3,2,2)m_{c}-6\hat{I}_0^{[1,0]}(3,2,1)m_{c}-2\hat{I}_{1}(3,1,1)m_{c}+8\hat{I}_1^{[1,0]}(2,1,3)m_{s}
\\
&&+4\hat{I}_2^{[1,0]}(3,2,1)m_{s}+12\hat{I}_{1}(1,1,3)m_{s}-4\hat{I}_{2}(1,2,2)m_{s}+2\hat{I}_2^{[0,1]}(2,2,2)m_{s}
\\
&&-2\hat{I}_{2}(2,2,1)m_{s}+4\hat{I}_2^{[0,1]}(3,1,2)m_{s}-2\hat{I}_2^{[1,1]}(3,2,2)m_{s}+4\hat{I}_{1}(2,1,2)m_{s}
\\
&&+4\hat{I}_{2}(1,1,3)m_{s}+4\hat{I}_{0}(2,1,2)m_{s}-10\hat{I}_{2}(2,1,2)m_{s}+2\hat{I}_2^{[1,0]}(2,2,2)m_{s}
\\
&&+4\hat{I}_2^{[0,1]}(2,1,3)m_{s}+20\hat{I}_{0}(1,1,3)m_{s},
\end{eqnarray*}
\begin{eqnarray*}
C_{D^*_{s0}D_1K^*}^{K^*}&=&2\hat{I}_{1}(3,2,2)m_{c}^{5}+2\hat{I}_{2}(3,2,2)m_{c}^{5}+2\hat{I}_{0}(3,2,2)m_{c}^{5}+2\hat{I}_{2}(3,2,2)m_{c}^{4}m_{s}
\\
&&-2\hat{I}_{2}(3,2,2)m_{c}^{3}m_{s}^{2}-2\hat{I}_{1}(3,2,2)m_{c}^{3}m_{s}^{2}-2\hat{I}_{0}(3,2,2)m_{c}^{3}m_{s}^{2}-2\hat{I}_{2}(3,2,2)m_{c}^{2}m_{s}^{3}
\\
&&+2\hat{I}_1^{[1,0]}(3,2,2)m_{c}^{3}-2\hat{I}_{2}(3,2,1)m_{c}^{3}+2\hat{I}_2^{[0,1]}(3,2,2)m_{c}^{3}+2\hat{I}_0^{[0,1]}(3,2,2)m_{c}^{3}
\\
&&+4\hat{I}_{1}(2,2,2)m_{c}^{3}+2\hat{I}_0^{[1,0]}(3,2,2)m_{c}^{3}+4\hat{I}_{0}(2,2,2)m_{c}^{3}+6\hat{I}_{2}(4,1,1)m_{c}^{3}
\\
&&+2\hat{I}_2^{[1,0]}(3,2,2)m_{c}^{3}+4\hat{I}_{2}(2,2,2)m_{c}^{3}+2\hat{I}_{2}(3,1,2)m_{c}^{3}+2\hat{I}_{0}(3,1,2)m_{c}^{3}
\\
&&+6\hat{I}_{1}(4,1,1)m_{c}^{3}+2(\hat{I}_0^{[0,1]}(3,2,2)m_{c}^{3}+6\hat{I}_{0}(4,1,1)m_{c}^{3}+2\hat{I}_2^{[0,1]}(3,2,2)m_{c}^{2}m_{s}
\\
&&+6\hat{I}_{2}(4,1,1)m_{c}^{2}m_{s}-2\hat{I}_{1}(2,2,2)m_{c}^{2}m_{s}+2\hat{I}_{0}(3,1,2)m_{c}^{2}m_{s}-2\hat{I}_{1}(3,1,2)m_{c}^{2}m_{s}
\\
&&+8\hat{I}_{0}(2,1,3)m_{c}^{2}m_{s}+2\hat{I}_{2}(3,1,2)m_{c}^{2}m_{s}+8\hat{I}_{1}(2,1,3)m_{c}^{2}m_{s}+2\hat{I}_{2}(2,2,2)m_{c}^{2}m_{s}
\\
&&+2\hat{I}_2^{[1,0]}(3,2,2)m_{c}^{2}m_{s}+4\hat{I}_{2}(2,1,3)m_{c}^{2}m_{s}-2\hat{I}_{0}(2,2,2)m_{c}^{2}m_{s}-2\hat{I}_1^{[1,0]}(3,2,2)m_{c}m_{s}^{2}
\\
&&+12\hat{I}_{1}(1,1,4)m_{c}m_{s}^{2}+6\hat{I}_{0}(3,1,2)m_{c}m_{s}^{2}+4\hat{I}_{1}(3,1,2)m_{c}m_{s}^{2}+12\hat{I}_{2}(1,1,4)m_{c}m_{s}^{2}
\\
&&+2\hat{I}_{2}(3,2,1)m_{c}m_{s}^{2}-2\hat{I}_0^{[1,0]}(3,2,2)m_{c}m_{s}^{2}-2\hat{I}_2^{[1,0]}(3,2,2)m_{c}m_{s}^{2}+12\hat{I}_{0}(1,1,4)m_{c}m_{s}^{2}
\\
&&+4\hat{I}_{2}(3,1,2)m_{c}m_{s}^{2}-2\hat{I}_2^{[1,0]}(3,2,2)m_{s}^{3}+4\hat{I}_{2}(3,1,2)m_{s}^{3}+12\hat{I}_{2}(1,1,4)m_{s}^{3}
\\
&&+4\hat{I}_{2}(2,2,2)m_{s}^{3}-4\hat{I}_{2}(2,1,3)m_{s}^{3}-6\hat{I}_2^{[0,1]}(3,1,2)m_{c}-2\hat{I}_{2}(2,2,1)m_{c}
\\
&&+4\hat{I}_{0}(1,2,2)m_{c}-4\hat{I}_1^{[1,0]}(3,2,1)m_{c}+4\hat{I}_{2}(1,2,2)m_{c}+10\hat{I}_{0}(2,1,2)m_{c}
\\
&&+2\hat{I}_1^{[1,1]}(3,2,2)m_{c}+4\hat{I}_{1}(1,2,2)m_{c}+8\hat{I}_{1}(2,1,2)m_{c}+10\hat{I}_{2}(2,1,2)m_{c}
\\
&&-4\hat{I}_2^{[1,0]}(3,2,1)m_{c}-4\hat{I}_1^{[0,1]}(3,1,2)m_{c}-6\hat{I}_0^{[0,1]}(3,1,2)m_{c}+2\hat{I}_{0}(3,1,1)m_{c}
\\
&&+2\hat{I}_0^{[1,1]}(3,2,2)m_{c}-2\hat{I}_2^{[0,1]}(3,2,1)m_{c}-4\hat{I}_0^{[1,0]}(3,2,1)m_{c}+2\hat{I}_2^{[1,1]}(3,2,2)m_{c}
\\
&&+2\hat{I}_{2}(1,2,2)m_{s}-4\hat{I}_2^{[1,0]}(3,2,1)m_{s}-2\hat{I}_{1}(1,2,2)m_{s}+2\hat{I}_0^{[1,0]}(2,2,2)m_{s}
\\
&&+6\hat{I}_{1}(1,3,1)m_{s}+4\hat{I}_{2}(1,1,3)m_{s}-2\hat{I}_2^{[1,0]}(2,2,2)m_{s}-4\hat{I}_{1}(1,1,3)m_{s}
\\
&&-2\hat{I}_{1}(2,1,2)m_{s}-4\hat{I}_2^{[0,1]}(2,2,2)m_{s}-6\hat{I}_{0}(2,1,2)m_{s}-2\hat{I}_{0}(1,2,2)m_{s}
\\
&&+4\hat{I}_2^{[0,1]}(2,1,3)m_{s}-4\hat{I}_{0}(1,1,3)m_{s}+2\hat{I}_1^{[1,0]}(2,2,2)m_{s}-4\hat{I}_2^{[0,1]}(3,1,2)m_{s}
\\
&&+2\hat{I}_2^{[1,1]}(3,2,2)m_{s}+2\hat{I}_{2}(2,1,2)m_{s}+4\hat{I}_{2}(2,2,1)m_{s},
\end{eqnarray*}

where
\begin{eqnarray*}
\hat{I}_{\mu}^{[\alpha,\beta]} (a,b,c)&=&
[M_1^2]^{\alpha}[M_2^2]^{\beta}\frac{d^\alpha}{d(M_1^2)^{\alpha}}
\frac{d^\beta}{d(M_2^2)^{\beta}}[M_1^2]^{\alpha}[M_2^2]^{\beta}\hat I_{\mu} (a,b,c),
\nonumber \\ \hat{I}_k(a,b,c) \!\!\! &=& \!\!\! i
\frac{(-1)^{a+b+c+1}}{16 \pi^2\,\Gamma(a) \Gamma(b) \Gamma(c)}
(M_1^2)^{1-a-b+k} (M_2^2)^{4-a-c-k} \, {U}_0(a+b+c-5,1-c-b),
\nonumber \\ \hat{I}_m(a,b,c) \!\!\! &=& \!\!\! i
\frac{(-1)^{a+b+c+1}}{16 \pi^2\,\Gamma(a) \Gamma(b) \Gamma(c)}
(M_1^2)^{-a-b-1+m} (M_2^2)^{7-a-c-m} \, {U}_0(a+b+c-5,1-c-b),
\nonumber\\ \hat{I}_6(a,b,c) \!\!\! &=& \!\!\! i
\frac{(-1)^{a+b+c+1}}{32 \pi^2\,\Gamma(a) \Gamma(b) \Gamma(c)}
(M_1^2)^{3-a-b} (M_2^2)^{3-a-c} \, {U}_0(a+b+c-6,2-c-b), \nonumber\\
\hat{I}_n(a,b,c) \!\!\! &=& \!\!\! i \frac{(-1)^{a+b+c}}{32
\pi^2\,\Gamma(a) \Gamma(b) \Gamma(c)} (M_1^2)^{-4-a-b+n}
(M_2^2)^{11-a-c-n} \, {U}_0(a+b+c-7,2-c-b),
\end{eqnarray*}
where $k=1, 2$, $m=3, 4, 5$ and $n=7, 8$. We can define the function
$U_0(a,b)$ as:
\begin{eqnarray*}
U_0 (a, b) = \int_0^{\infty} dy (y + M_1^2 + M_2^2)^ay^b \exp
[-\frac{B_{-1}}{y} - B_0 - B_1y ],
\end{eqnarray*}
where
\begin{eqnarray*}
B_{-1} &=& \frac{1}{M_2^2M_1^2}(m_s^2(M_1^2 +M_2^2)^2
-M_2^2M_1^2Q^2), \nonumber\\
B_{0} &=& \frac{1}{M_1^2M_2^2}(m_s^2 + m_c^2)(M_1^2+M_2^2)
,\\
B_{1} &=& \frac{m_c^2}{M_1^2M_2^2}. \nonumber
\end{eqnarray*}

\end{document}